\documentclass[aps,twocolumn,superscriptaddress,nofootinbib,a4paper,longbibliography]{revtex4-2}

\usepackage{times}
\usepackage{epsfig}
\usepackage{amsfonts}
\usepackage{amsmath}
\usepackage{amsthm}
\usepackage{amssymb}					
\usepackage{amsthm}
\usepackage{dsfont}
\usepackage{bm}
\usepackage{mathtools}
\usepackage{color}
\usepackage{multirow}
\usepackage[normalem]{ulem}
\newcommand{\stkout}[1]{\ifmmode\text{\sout{\ensuremath{#1}}}\else\sout{#1}\fi}
\usepackage{latexsym}
\usepackage{mathrsfs}
\usepackage{natbib}
\usepackage{verbatim}
\usepackage[T1]{fontenc}
\usepackage{float}
\usepackage{graphicx}
\usepackage{xcolor}
\usepackage{physics}
\usepackage{soul}
\usepackage{rotating}
\usepackage{tikz}
\usetikzlibrary{tikzmark}

\theoremstyle{definition}

\newcommand{\bracket}[3]{\langle#1|#2|#3\rangle}
\newcommand{\expect}[1]{\langle#1\rangle}

\usepackage[colorlinks=true,linkcolor=blue,citecolor=magenta,urlcolor=blue]{hyperref}

\begin{document}


\title{Imprecision plateaus in quantum steering}

\author{Elna Svegborn}
\affiliation{Physics Department and NanoLund, Lund University, Box 118, 22100 Lund, Sweden.}

\author{Nicola d'Alessandro}
\affiliation{Physics Department and NanoLund, Lund University, Box 118, 22100 Lund, Sweden.}

\author{Otfried G\"uhne}
\affiliation{Naturwissenschaftlich-Technische Fakult\"at, Universit\"at Siegen, Walter-Flex-Stra{\ss}e 3, 57068 Siegen, Germany}

\author{Armin Tavakoli}
\affiliation{Physics Department and NanoLund, Lund University, Box 118, 22100 Lund, Sweden.}

\begin{abstract}
We study tests of quantum steering in which the trusted party does not have perfect control of their measurements. We report on steering inequalities that remain unaffected when introducing up to a threshold magnitude of measurement imprecision. This phenomenon, which we call an imprecision plateau, thus permits a departure from the standard assumption of idealised measurements without any incuring cost to the detection power of steering experiments. We provide an explanation for why imprecision plateaus are possible, a simple criterion for their existence and tools for analysing their properties. We also demonstrate that these plateaus have natural applications when the assumption of perfect measurements is relaxed: they allow for maintaining both the noise- and loss-robustness of standard steering tests and the performance rate of idealised one-sided device-independent random number generators. 
\end{abstract}

\date{\today}

\maketitle

\textit{Introduction.---} Quantum steering is the impossibility for a hidden variable explaining how quantum measurements on one particle can alter the local state of another particle.  The scenario can be understood as a black-box party, Alice, influencing the lab of another party, Bob, in a way that proves shared entanglement. The last two decades have seen much interest in studying quantum steering in theory and experiments \cite{Uola2020, Cavalcanti2017}, and developing its applications within e.g. one-sided device-independent quantum key distribution \cite{Branciard_2012, ma2011}, randomness certification \cite{Law_2014, Li2024} and quantum teleportation \cite{He2015}. 

In the steering scenario, Alice selects an input, $x$, and performs a corresponding measurement, $\{A_{a|x}\}_a$, on her share of a bipartite state $\rho_{AB}$. The (sub-normalised) states created remotely for Bob become $\sigma_{a|x}=\tr_A\left(A_{a|x}\otimes \openone \rho_{AB}\right)$. The set $\{\sigma_{a|x}\}$  is called the steering assemblage. Bob must now determine whether the assemblage can be modelled via a local hidden state (LHS), i.e.~whether it admits the form $\sigma_{a|x}=\sum_\lambda p(\lambda)p(a|x,\lambda)\sigma_\lambda$, where $\lambda$ is a hidden variable and $\sigma_\lambda$ is the hidden quantum state \cite{Wiseman2007}. If the answer is negative, the assemblage is steerable, implying that  $\rho_{AB}$ is entangled. To detect steerability, Bob must select an input $y$ and perform an associated measurement $\{B_{b|y}^\text{targ}\}_b$. The assemblage demonstrates steering if and only if the input/output statistics can violate a suitable inequality that is respected by all local hidden state (LHS) models but not by quantum theory \cite{Wiseman2007}. Such a steering inequality takes the form 
\begin{equation}\label{steerineq}
\mathcal{W}_0\equiv \sum_{a,b,x,y}  c_{abxy}\tr(\sigma_{a|x} B^\text{targ}_{b|y}) \stackrel{\text{LHS}}{\leq} \beta_0,
\end{equation}
where $c_{abxy}$ are arbitrary real coefficients and $\beta_0$ is the tight bound for LHS models.



\begin{figure}
	\centering
	\includegraphics[width=\columnwidth]{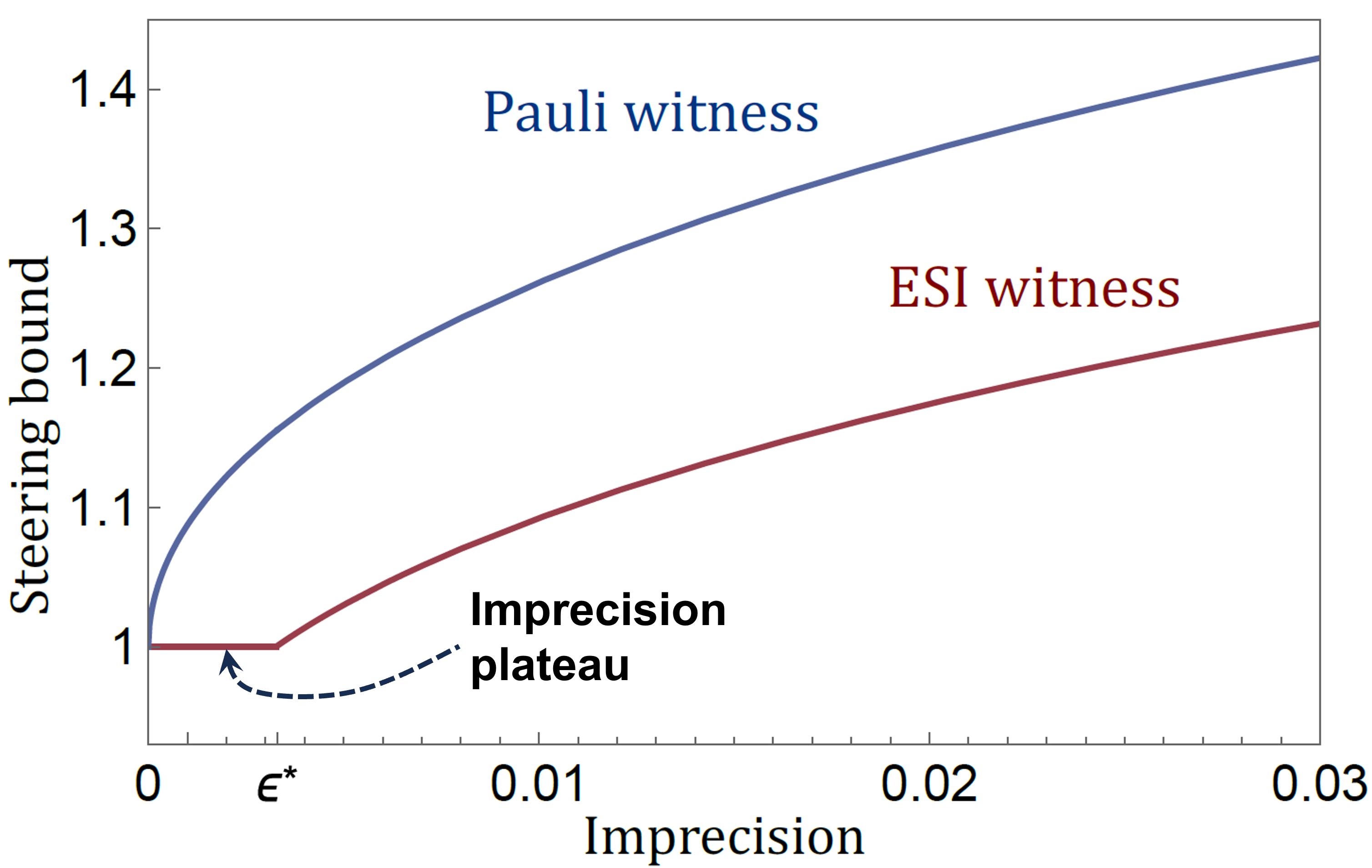}
	\caption{Imprecise measurements for the trusted party typically leads to stronger LHS correlations, i.e.~a higher steering bound (blue curve). However, some steering witnesses remain unchanged, up to a critical value, even though the measurements are increasingly uncontrolled. This leads to an imprecision plateau (red curve).}\label{Fig_plateau}
\end{figure}

However, standard steering inequalities \eqref{steerineq} require a strong assumption: that Bob is able to \textit{perfectly} implement his targeted measurements. This is an idealisation to which experiments can only aspire. A series of works, focused on experiments for tomography \cite{Rosset2012}, communication \cite{Tavakoli2021, AlmostQudit}, entanglement \cite{Moroder2012, Rosset2012, Morelli2022}, and steering \cite{Moroder2016, Tavakoli2024, Sarkar2023}, have shown that if Bob's operations only nearly correspond to those targeted, it can lead to a high risk of false positives. For example, the perhaps most well-known test of two-qubit steering involves Alice and Bob measuring the three Pauli observables $\{X,Y,Z\}$. Specifically, in an LHS model, this ``Pauli witness'' is 
\begin{equation}\label{eq:Pauli_witness}
\mathcal{W}_0^\text{pauli}=\frac{1}{\sqrt{3}}\left(\expect{A_1 \otimes X}-\expect{A_2 \otimes Y}+\expect{A_3 \otimes Z}\right)\leq 1
\end{equation}
for arbitrary observables $A_i$ with eigenvalues $\pm 1$, where $\expect{\cdot}$ denotes the expectation value \cite{Cavalcanti2009, Saunders2010}. Quantum theory achieves violations up to $\mathcal{W}_0^\text{pauli}=\sqrt{3}$. However, if Bob's lab observables turn out to have only a nearly perfect fidelity of, say, $99.5\%$, over $25\%$ of the possible quantum violation range can be explained by false positives stemming from LHS models \cite{Tavakoli2024}. The high-sensitivity to small measurement imprecisions is illustrated for the Pauli witness by the steep increase around the origin of the blue curve in Fig.~\ref{Fig_plateau}. Similar effects arise for other common steering inequalities \cite{Tavakoli2024}, and only some refined methods are known that can detect steering without this drawback \cite{Moroder2016}.

 
It has been proposed to eliminate the perfect-measurement assumption by replacing it with a quantitative limitation of Bob's lack of control. Refs.~\cite{Rosset2012, Cao2023, Morelli2022, Tavakoli2024} consider the case where Bob aims to perform a target measurement corresponding to a basis $\{\ket{e_b}\}$, but instead implements a different lab measurement  $\{E_b\}$. The lab measurement can be subject to arbitrary imperfection; it is only assumed to be a good approximation of the target measurement, fulfilling the constraint
\begin{equation}\label{imprecise}
\bracket{e_b}{E_b}{e_b}\geq 1-\epsilon_b,
\end{equation} 
for some small imprecision parameter $\epsilon_b\in[0,1]$. This definition has the advantages of being platform-independent and operationally meaningful. Indeed, $\epsilon_b$ can be estimated with small resources in the lab. However, when permitting imprecise measurements in our model, we reduce our assumed prior knowledge of the experiment. This means that the LHS bound in \eqref{steerineq} in general must increase to account for all measurements compatible with the condition \eqref{imprecise}, leading to an $\epsilon$-dependent criterion, $\mathcal{W}_\epsilon\leq \beta_\epsilon$ (e.g.,~the blue curve in Fig.~\ref{Fig_plateau}).

Here, we reveal that it is possible to conduct steering tests without assuming idealised measurements that nevertheless cost no additional detection power. Specifically, we show that there exists steering inequalities of the form \eqref{steerineq} where the LHS bound does not change when relaxing the assumption of perfectly controlled measurements up to a critical imprecision magnitude $\epsilon^*$. That is, when the imprecision is below the threshold, i.e.~$\epsilon\leq \epsilon^*$, the steering inequality remains  $\mathcal{W}_\epsilon\leq \beta_0$. This corresponds to an \textit{imprecision plateau} for steering inequalities, in which the ability to detect steering remains as good as if the measurements had been assumed perfect (see red curve in Fig.~\ref{Fig_plateau}).  We give an intuitive explanation for the existence of imprecision plateaus in terms of properties of well-chosen steering witnesses. For our main example, we analytically characterise the plateau's length (see $\epsilon^*$ in Fig.~\ref{Fig_plateau}), while for generic steering witnesses we develop both analytical and numerical relaxation methods to estimate it. We then proceed to showcase the utility of these plateaus in tests of steering and its quantum information applications.

\textit{Imprecision plateaus.---} To show the existence of imprecision plateaus, we consider a steering witness based on the so-called Elegant Bell Inequality \cite{Gisin2009, Tavakoli2020platonicsolids}, which we, in analogy, will refer to as the Elegant Steering Inequality (ESI). In the ESI scenario, Alice and Bob have four and three inputs, respectively, and binary outputs. Alice's measurements are unknown, whereas Bob's three target measurements correspond to the Paulis, i.e.,  $B_1^\text{targ}=X$, $B_2^\text{targ}=Y$, and $B_3^\text{targ}=Z$. The ESI reads 
\begin{align}\nonumber\label{EBI}
\mathcal{W}_0^\text{ESI}&=\frac{1}{4}\Big(\expect{\left(A_1+A_2-A_3-A_4\right)\otimes B_1^\text{targ}}\\\nonumber
&+\expect{\left(A_1-A_2+A_3-A_4\right)\otimes B_2^\text{targ}}\\
&+\expect{\left(A_1-A_2-A_3+A_4\right)\otimes B_3^\text{targ}}\Big)\stackrel{\text{LHS}}{\leq} 1.
\end{align}
\noindent Here, Alice's observables are defined as $A_x= A_{0\mid x} - A_{1 \mid x}$ and can be conveniently written as $A_x=\vec{n}_x\cdot (X,Y,Z)$ for each input $x = \{1,2,3,4\}$, where  $\vec{n}_x$ are Bloch vectors. The  LHS bound can be computed as a largest-eigenvalue problem. The maximal quantum violation of the ESI is $\mathcal{W}_0=\sqrt{3}$ and it is reached by sharing a maximally entangled state $\ket{\phi^+}=(\ket{00}+\ket{11})/\sqrt{2}$ and selecting $\{\vec{n}_x\}_{x=1}^4$ as $\{(1,-1,1),(1,1,-1),(-1,-1,-1),(-1,1,1)\}/\sqrt{3}$, forming a regular tetrahedron on the Bloch sphere. This choice gives $\mathcal{W}_0^\text{ESI}=(\expect{X\otimes X}-\expect{Y\otimes Y}+\expect{Z\otimes Z})/\sqrt{3}$ which is identical to the optimal way to violate the Pauli witness \eqref{eq:Pauli_witness}.


Now, we replace the target measurements in \eqref{EBI} with $\epsilon$-imprecise measurements, $\{B_y^\epsilon\}$, and set out to compute the new, $\epsilon$-dependent, LHS bound. From Eq.~\eqref{imprecise}, one finds that this means considering all qubit observables such that $\tr(B_y^\text{targ}B_y^\epsilon)\geq 2-4\epsilon$. Since the witness is linear, we need only to consider all deterministic input-output strategies of Alice. 
Labelling them by $(a_1,a_2,a_3,a_4)\in\{0,1\}^4$, the values of $\mathcal{W}_\epsilon$ become \cite{footnote1} 
\begin{align}\nonumber\label{strats}
&\quad(0,0,0,0) \Rightarrow \mathcal{W}_\epsilon=0\\[5pt]\nonumber
&\quad(0,0,0,1) \Rightarrow \mathcal{W}_\epsilon= \expect{B_1^\epsilon+B_2^\epsilon-B_3^\epsilon}_\psi /2 && \\ 
&\quad(0,0,1,0) \Rightarrow \mathcal{W}_\epsilon= \expect{B_1^\epsilon-B_2^\epsilon+B_3^\epsilon}_\psi /2 &&\\\nonumber
&\quad(1,0,1,1) \Rightarrow \mathcal{W}_\epsilon= \expect{B_1^\epsilon-B_2^\epsilon-B_3^\epsilon}_\psi /2 &&\\\nonumber
&\quad(0,1,1,1) \Rightarrow \mathcal{W}_\epsilon=\expect{B_1^\epsilon+B_2^\epsilon+B_3^\epsilon}_\psi /2 &&=f^*(\epsilon)\\[5pt]\nonumber
&\quad(0,0,1,1) \Rightarrow \mathcal{W}_\epsilon=\expect{B_1^\epsilon}_\psi&&\\\nonumber
&\quad(0,1,0,1) \Rightarrow \mathcal{W}_\epsilon=\expect{B_2^\epsilon}_\psi&&\\\nonumber
&\quad(0,1,1,0) \Rightarrow \mathcal{W}_\epsilon=\expect{B_3^\epsilon}_\psi&&=g^*(\epsilon).
\end{align}
\begin{tikzpicture}[remember picture, overlay]
	\draw[thick] (0.51,4.55)--(0.51,4.95);
	\node at (0.3, 4.75) {\rotatebox{90}{$C1$}};
	\draw[thick] (0.51,2.4)--(0.51,4.2);
	\node at (0.3, 3.4) {\rotatebox{90}{$C2$}};
	\draw[thick] (0.51,0.7)--(0.51,2);
	\node at (0.3, 1.5) {\rotatebox{90}{$C3$}};
\end{tikzpicture}

Due to the Bloch sphere symmetry and the unbiasedness of the three target measurements, we see that these eight cases  reduce to just  three  qualitatively different classes, as indicated above. Class 1 is trivial. For class 2, the steering bound is $
f(\epsilon)\equiv \max f^*(\epsilon)=\dfrac{\sqrt{3}}{2}\left(1+2\sqrt{2\epsilon(1-\epsilon)}-2\epsilon\right)$, 
which is proven in Supplementary Material (SM) I. Note that for $\epsilon=0$ we have $f=\frac{\sqrt{3}}{2}$. Next,  we make the key observation: the value for class 3 is \textit{independent} of $\epsilon$, because trivially $g(\epsilon)\equiv\max g^*(\epsilon)=1$. Since $g(0)>f(0)$, this is the LHS bound appearing in \eqref{EBI}. Crucially, the bound $\mathcal{W}_\epsilon^\text{ESI}\leq 1$ remains valid until $\epsilon$ is large enough for $f(\epsilon)$ to overtake $g(\epsilon)=1$. This happens at  
\begin{equation}
\epsilon^*=\frac{1}{2}-\frac{1}{3\sqrt{3}}-\frac{1}{3}\sqrt{\frac{5}{6}}\approx 3.3\times 10^{-3}.
\end{equation}
Thus, the ESI witness under imprecise measurements becomes $\mathcal{W}_\epsilon^\text{ESI}\leq 1$ for $\epsilon\leq \epsilon^*$ and $\mathcal{W}_\epsilon^\text{ESI}\leq f(\epsilon)$ for $\epsilon^*< \epsilon \leq (3-\sqrt{3})/6$, where the upper demarcation corresponds to the maximal value of $f$ \cite{footnote2}. This threshold implies the imprecision plateau (see red curve in Fig.~\ref{Fig_plateau}) and $\epsilon^*$ can therefore be interpreted as the plateau's length.

Importantly, imprecision plateaus as a phenomenon are \textit{independent} of the specific quantifier of measurement imprecision. For instance, we could replace the fidelity measure \eqref{imprecise} with a limitation on the norm of the anti-commutators $\|\{B_y,B_{y'}\}\|_\infty \leq \varepsilon$  \cite{Gois2023}. We still have $g(\varepsilon)=1$ for class 3 and via the Cauchy-Schwarz inequality we obtain for class 2,
$f^*(\epsilon)\leq \sqrt{ 3+\smash{\sum_{j<k}}\expect{\{B^{\smash{\varepsilon}}_{j},B^{\smash{\varepsilon}}_{k}\} }_{\smash{\psi}}}/2 \leq \sqrt{3(1+\varepsilon)}/2 =f(\varepsilon)$. For $\varepsilon\leq \frac{1}{3}$, $f(\varepsilon)$ does not exceed $g(\varepsilon)=1$ and hence, we have a plateau in $\varepsilon$.

\textit{Construction of plateau witnesses.---} In general, a steering witness supports an imprecision plateau if and only if all deterministic strategies associated with the tight LHS bound ($\epsilon=0$) correspond to witness expressions [as in Eq.~\eqref{strats}] that depend only on a \textit{single} input of Bob. In an LHS model, this permits Bob's local state to align itself with the relevant observable independently of $\epsilon$. In contrast, this is impossible when the witness expression depends on more than a single observable, because the metric distance of two imprecise observables will be $\epsilon$-dependent. This allows us to determine whether a given witness has an imprecision plateau by solving a set of largest-eigenvalue problems. 

However, how does one construct witnesses that have this property? A simple systematic way is to take $\mathcal{W}_0=\sum_{x,y}c_{xy}\expect{A_x\otimes B^\text{targ}_y}$ and select the coefficient matrix, $c$, with the following properties:  (i) the first column is  $(1,\ldots,1)^T$ and (ii) all other columns individually sum to zero  \cite{Footnote}. By construction, if Alice always outputs $A_x=1$ then $\mathcal{W}_0\propto B^\text{targ}_1$. Other deterministic strategies may correspond to expectation values involving several observables. Since the LHS value associated with the first type of strategy is independent of the choice of $B^\text{targ}$, whereas the other are not, one can then select $\{B^\text{targ}_y\}$ so that the former corresponds to the LHS bound. This implies the existence of an imprecision plateau.

We now construct a relevant family of plateau witnesses that naturally generalise the ESI. The family is parameterised by an integer $n\geq 3$. Let Alice have $2^{n-1}$ possible inputs represented as a string $x=(0,x_2,\ldots,x_{2^{n-1}})$ with $x_i\in\{0,1\}$. Let Bob have $n$ target observables, selected so that they pairwise anti-commute i.e.~$\{B^\text{targ}_j,B^\text{targ}_{k}\}=0$ for $j\neq k$. Next, define witness coefficients $c_{xy} = (-1)^{x_y}/2^{n-1}$ and note that properties  (i,ii) are satisfied. Note also that for $n=3$ this is equivalent to the ESI witness. In SM II we first prove that the LHS bound for this witness is $\beta_0=1$ $\forall n$ and then we prove that all the deterministic strategies that saturate it correspond to witness expressions that depend only on one target observable . Thus, this witness has a plateau for every $n$. These witnesses has two noteworthy properties. (I) The steering inequality is valid in any Hilbert space dimension. In particular, a set of $n$ anti-commuting target observables can always be identified in dimension $d=2^{\lfloor n/2\rfloor}$. Thus, by selecting $n$, this witness family is tailored for steering in arbitrary high-dimensional Hilbert spaces. (II) A quantum violation exists achieving $\mathcal{W}_0=\sqrt{n}$ by using a $d$-dimensional maximally entangled state (see SM II). Consequently, even though the noise tolerance improves with $n$ the imprecision plateau always remains possible.

\textit{Semidefinite programming methods.---} While the ESI is simple enough for analytically determining the plateau's length, this is in general rarely the case. Consider that we are given an arbitrary steering witness and we find that it supports a plateau. How can we bound the plateau's length? Already for the more experimentally relevant variant of the ESI, in which each of Bob's three measurements has a different imprecision parameter, $\vec{\epsilon}=(\epsilon_X,\epsilon_Y,\epsilon_Z)$, we have been unable to solve the problem exactly. One possibility is to apply the main theorem of Ref.~\cite{Tavakoli2024} to each of the deterministic strategies of Alice. This theorem gives a bound on the operators in terms of the imprecision parameter and  allows one to compute $\epsilon$-dependent upper bounds on any steering witness $\mathcal{W}_\epsilon$ as perturbations to the standard steering inequality \eqref{steerineq}, but it does not give a tight bound (see SM III). Another, less rigorous option, also discussed in SM III, is to use a Taylor approximation in the relevant regime of small imprecision. This gives better bounds, but still not tight. To overcome this, we instead address the problem of bounding the plateau length using semidefinite programming (SDP) relaxations based on moment matrices \cite{TavakoliSDP}.

The method, discussed in detail in SM IV, is based on three main ideas. Firstly, contrasting the common approach to Bell-type scenarios \cite{Navascues2007, Pusey2013}, we define a tracial moment matrix, $\Gamma_{u,v}=\tr(u^\dagger v)$, where $u$ and $v$ are monomials of observables in the selected relaxation. Secondly, we use the idea of \cite{Navascues2015} to add dimensionality constraints on $\Gamma$ by constructing a basis of moment matrices via random sampling. Thirdly, we follow the idea of \cite{Tavakoli2021} and explicitly insert the target measurements into the monomial set to incorporate the fidelity constraints into the SDP relaxation. Combined, this lets us express the steering witness as a linear combination over $\Gamma$ and realise \eqref{imprecise} as linear constraints. Given that $\Gamma$ is positive semidefinite by construction, it can be solved as an SDP. To showcase the usefulness of this method, we have applied it to the ESI witness with the three independent imprecision parameters. Already at first level relaxation, corresponding to an SDP matrix of dimension 14 with 19 variables, it systematically returns bounds on the plateau length that match our brute-force numerical search values up to at worst five decimals. The plateau length in $\epsilon_Z$ as a function of $\epsilon_X$ and $\epsilon_Y$ is given in SM IV.

\textit{Application to noise and loss.---} Imprecision plateaus have natural applications in experiments that recognise the lack of perfect measurement control. Typically, as we have exemplified in Fig.~\ref{Fig_plateau} for the Pauli witness \eqref{eq:Pauli_witness}, the quantum violation gap rapidly shrinks when accounting for imperfect measurements. This is equivalent to the decrease of the corresponding noise tolerance of steering. However, the plateau permits experiments to account for measurement imperfections while  maintaining the noise-tolerance associated with idealised measurements. For example, for perfect measurements ($\epsilon=0$), both the Pauli witness and the ESI witness detect steering for the isotropic state $\rho_v=v\ketbra{\phi^+}+(1-v)\frac{\openone}{4}$ when $v>\frac{1}{\sqrt{3}}\approx 0.577$. However, when permitting measurement imprecision corresponding to the plateau length $\epsilon^*$, the equivalence is far gone. The former now requires $v\gtrsim 0.667$ whereas the latter still has $v\gtrsim 0.577$. Note that a nonlinear test for steering exists, where Bob implements Pauli measurements on qubits without any assumption
on the precision \cite{Moroder2016}. This works for any $v\gtrsim 0.577$ too, but is not straightforward to generalize to arbitrary systems.

A similar advantage applies for scenarios with detection losses, i.e., when Alice's measurement fails with probability $\eta\in[0,1]$.  Since in our picture the measurement errors of Bob are uncontrolled, we do not allow him to actively use the $\epsilon$-cone around his target measurement toward closing the detection loophole. Instead, we assume that he performs the target measurements. It is well-known that the critical efficiency, $\eta_\text{crit}$, needed for demonstrating steering is much lower than the efficiency needed for a Bell inequality violation, where Bob's measurements are entirely uncharacterised \cite{Bennet2012, Wittmann2012, Evans2013}. Therefore, with the gradual loss of measurement control, one expects $\eta_\text{crit}$ to increase towards the Bell inequality limit. However, the imprecision plateau permits us to circumvent this problem for imprecision magnitudes within the plateau. For perfect measurement control, the critical detection efficiency for both the Pauli witness and the ESI witness is $\eta_\text{crit}=\frac{1}{3}$. However, at imprecision $\epsilon=\epsilon^*$, the advantage becomes large: the ESI witness still has $\eta_\text{crit}=\frac{1}{3}$ whereas the Pauli witness now requires $\eta_\text{crit}\approx  0.606$. In SM V, we compute the optimal witness values at fixed detection efficiency.

\textit{Application: random number generation.---} Imprecision plateaus also enable advantages in one-sided device-independent protocols. We showcase this for a quantum random number generation scenario, in which we restrict ourselves to  individual eavesdropper attacks, see e.g. Ref.~\cite{Tomamichel2012} for more general types of attacks. Using the ESI witness as a testbed, we show that for imprecisions within the plateau, one can both certify randomness whenever the original steering inequality is violated and obtain the same rate of randomness as when assuming perfect measurements. Thus, the performance of the protocol is fully maintained while eliminating idealisations that otherwise may be used to compromise the security.

\begin{figure}[t!]
	\centering
	\includegraphics[width=\columnwidth]{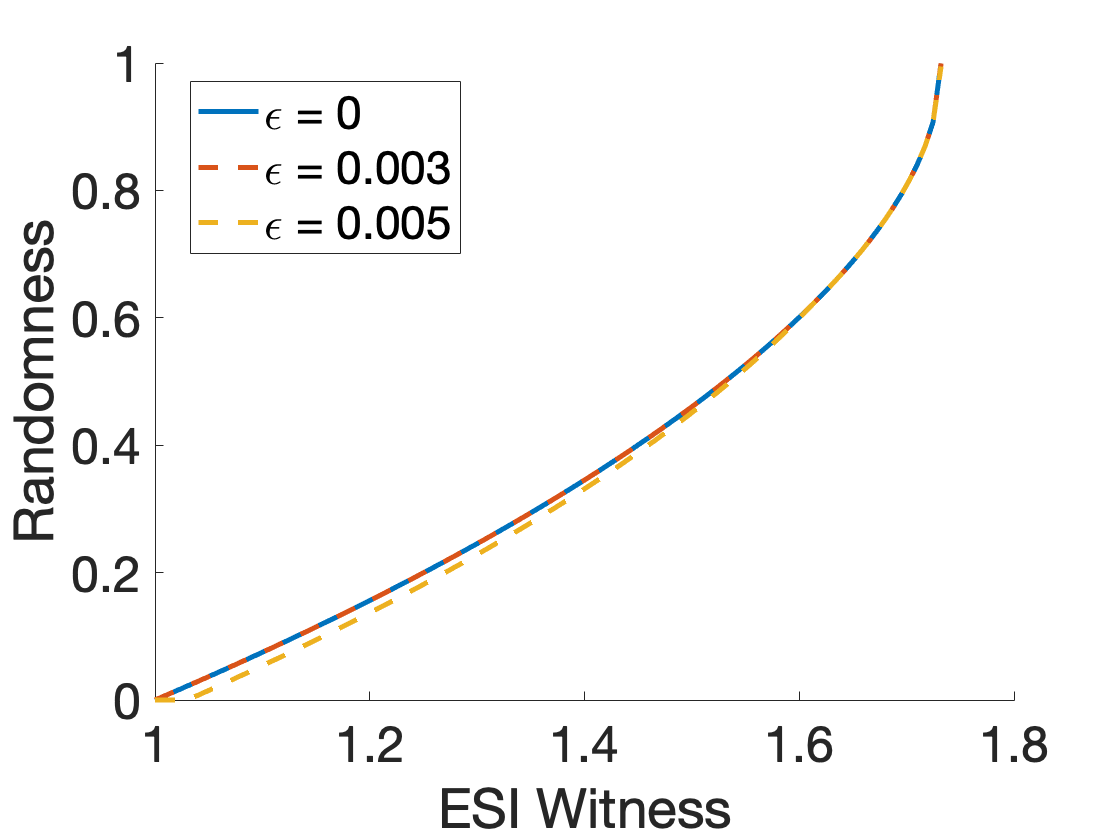}
	\caption{The performance of the randomness generation remains unchanged when introducing measurement imprecisions below the plateau length (blue and red curve). However, imprecisions above the plateau length lead to the expected decrease in noise tolerance and generation rate (yellow curve).}\label{Fig_rand}
\end{figure}

We certify randomness from Alice's output averaged over her four possible inputs. The randomness is given by $R=-\log_2(P_g)$ where $P_g=\max \frac{1}{4}\sum_{a,x}p(a|x)$ is the average probability of an eavesdropper guessing her outcome \cite{Pironio2010}, maximised over all strategies. Thus, we must compute the largest possible $P_g$ compatible with the observed value of the ESI witness. This can be achieved by evaluating $P_g$ separately for each of Alice's deterministic strategies. When $\epsilon=0$, each of these optimisations can be straightforwardly cast as an SDP. When $\epsilon>0$, we can bound $P_g$ from above (corresponding to a lower bound on $R$) by using the previously discussed SDP relaxation method. This is applied separately to each of the deterministic strategies \cite{Mironowicz2016}. To this end, we have selected $\epsilon=3\times 10^{-3}<\epsilon^*$ and employed a moment matrix of size 296 featuring 1371 variables. The randomness rates are illustrated in Fig.~\ref{Fig_rand}. Our certified randomness bound is essentially equal to the amount of randomness obtained for $\epsilon=0$, with a worst-case difference of  $\Delta R = 0.0016$. The small discrepancy is likely due to our choice of relaxation level; which is a compromise between accuracy and practicality. This illustrates the maintained protocol performance in face of measurement imprecisions. In contrast, when operating outside the plateau, e.g.~at $\epsilon=5\times 10^{-5}>\epsilon^*$, we cannot maintain neither the randomness rate nor the noise tolerance. This is illustrated by the yellow curve in Fig.~\ref{Fig_rand}, which is an upper bound on $R$ obtained from numerical search. It implies a clear reduction in the performance when not having access to the imprecision plateau.

\textit{Discussion.---}
Quantum steering experiments are commonly performed under the assumption that the trusted party has perfectly controlled measurements. This is an unrealistic assumption, whose near-validity is also rarely evidenced in practice. For steering scenarios, the problem of how to make witness tests valid under arbitrary imprecision was addressed in Ref.~\cite{Tavakoli2024}. However, the main drawback is that most of the well-known steering witnesses rapidly lose their detection power even as tiny imprecision is introduced. In this work, we have shown a sharp contrast to this observation, namely that well-chosen tests of steering can remain unchanged even when perfect control of the trusted measurement is relaxed. We have seen how this is made possible in steering because of the phenomenon of imprecision plateaus. These plateaus lead to several natural questions. How does one construct steering witnesses that both have a long imprecision plateau and detect entanglement that is far from Bell nonlocal? For a given steerable state, what is the longest imprecision plateau possible? It seems likely that imprecision plateaus also should appear more broadly, for example in multipartite steering or steering with quantum inputs \cite{Zhao2020,Kocsis2015}.

We have also shown that imprecision plateaus have natural applications. These can be divided into two types. Firstly, the plateaus pave the way for tests of steering that remove the undesirable assumption of perfect measurements, thus becoming more resilient to false positives, but without any cost to neither the noise or loss requirements of the experiment. We have exemplified these advantages in our analysis of the ESI.  Importantly, the plateau lengths that we have found are significantly larger than the imprecision parameters reported in a recent experiment  \cite{Cao2023}.  Secondly, they can be used to make one-sided device-independent quantum information protocols more faithful while maintaining the performance of the original, idealised, protocol. We illustrated this possibility in a proof-of-principle analysis of a quantum random number generator. This suggests that similar advantages may be more widely possible.

\begin{acknowledgements}
We thank Alexander Bernal, Elisa Monchietti, Carles Roch i Carceller, Owidiusz Makuta and Remigiusz Augusiak for discussions. This work is supported by the Wenner-Gren Foundations, by the Knut and Alice Wallenberg Foundation through the Wallenberg Center for Quantum Technology (WACQT), by the EDU-WACQT program funded by Marianne and Marcus Wallenberg Foundation, by the Swedish Research Council under Contract No. 2023-03498, by the Deutsche Forschungsgemeinschaft (DFG, German Research Foundation, project numbers 447948357 and 440958198, by the Sino-German Center for Research Promotion (Project M-0294), and the German Ministry of Education and Research (Project QuKuK, BMBF Grant No. 16KIS1618K).
\end{acknowledgements}

\bibliography{references_plateau_resub}

\begin{thebibliography}{39}
\bibitem [{foo({\natexlab{a}})}]{footnote1}%
  \BibitemOpen
  \href@noop {} {\bibinfo {title} {We need only to consider eight of {Alice's}
  16 deterministic strategies because those equivalent under a bit-flip
  correspond to an irrelevant global minus sign.}} ({\natexlab{a}})\BibitemShut
  {NoStop}%
\bibitem [{foo({\natexlab{b}})}]{footnote2}%
  \BibitemOpen
  \href@noop {} {\bibinfo {title} {The maximal value is $f=3/2$ and corresponds
  to the local hidden variable bound for the witness, i.e.~the {Elegant} {Bell}
  {Inequality}.}} ({\natexlab{b}})\BibitemShut {NoStop}%
\bibitem [{Foo()}]{Footnote}%
  \BibitemOpen
  \href@noop {} {\bibinfo {title} {Note also that one may flip the sign on any
  number of rows of $c$ without changing the plateau properties.}}\BibitemShut
  {Stop}%
\end{thebibliography}

\appendix

\section{Proof of the imprecision plateau of the ESI witness}\label{AppClass2}
We compute the witness value associated Alice's deterministic strategies, in the main text labelled as class 2, namely
\begin{equation}\label{eq:opt_ct}
f(\epsilon) \equiv \max_{\psi, B_y^\epsilon}  f^*(\epsilon) = \max_{\psi, B_y^\epsilon} \frac{1}{2}\expect{B_1^\epsilon+B_2^\epsilon+B_3^\epsilon}_\psi.
\end{equation}
This entails an optimisation over all states $\psi$ and qubit observables $B_y^\epsilon$, such that they are at least $\epsilon_y = \epsilon$ close to their associated target measurement $B_1^{\text{targ}} = X$, $B_2^{\text{targ}} = Y$, and $B_3^{\text{targ}} = Z$. We give two arguments for how to derive $f(\epsilon)$. The first is based on Bloch sphere geometry and the other is based on a strenghtened version of the main result of Ref.~\cite{Tavakoli2024}.

\subsection{Bloch sphere argument}
\begin{figure}
	\centering
	\includegraphics[width=0.9\columnwidth]{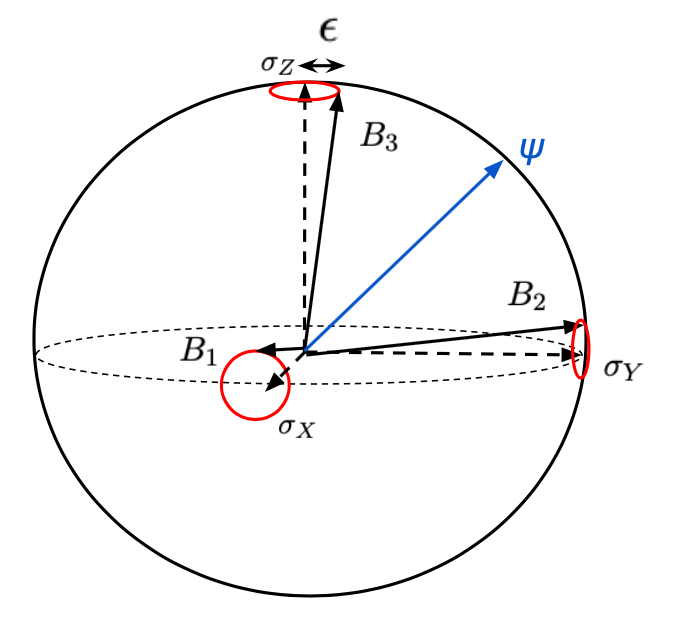}
	\caption{The imprecise measurements $B_y$ on the Bloch sphere can move in $\epsilon$-cones around the target measurements. We want the total overlap between $\psi$ and $B_y$ to be maximised.}\label{Fig_bloch_sphere}
\end{figure}

We start by translating the optimisation problem in Eq.~\eqref{eq:opt_ct} to the Bloch sphere, see schematic illustration in Fig.~\ref{Fig_bloch_sphere}. For perfectly controlled measurements, we find that $\max_{\psi} f^*(\epsilon = 0)$ reaches its maximum value $\sqrt{3}/2$ when $\psi$ is pure and overlaps equally with each of the three orthogonal observables, i.e., when its Bloch vector is given by $\vec{\psi} = (1,1,1)/\sqrt{3}$. This is obtained from a largest-eigenvalue calculation. In fact, because of the spherical symmetry of the Bloch sphere and the unbiasedness of $B_y^\epsilon$, we find that the maximal value of $f(\epsilon > 0)$ is obtained when $\psi$ remains in the middle of the sphere octant and $B_y^\epsilon$ move symmetrically towards it. Consequently, we can w.l.g. express the upper bound of $f(\epsilon)$ in terms of only a single observable, say $B_3^\epsilon$,
\begin{equation}
f(\epsilon) = \max_{B_y^\epsilon} \frac{1}{2}\langle B_1^\epsilon+B_2^\epsilon+B_3^\epsilon\rangle_\psi =  \dfrac{3}{2} \max_{B_3^\epsilon} \expect{B_3^\epsilon}_\psi.
\end{equation}
Next, we use that every extremal qubit observable $O$ can be expressed in terms of a vector on the Bloch sphere, i.e., $O = \vec{n} \cdot \vec{\sigma}$, where the normalised Bloch vector $\vec{n}$ is the direction of the measurement and $\vec{\sigma}  = (X,Y,Z)$. Thus, by associating $B_3^\epsilon$ with the Bloch vector $\vec{b} = (\sin\theta \cos\varphi, \sin\theta \sin\varphi, \cos\theta)$, where $\theta \in [0,\pi], \varphi \in [0,2\pi)$, we obtain that
\begin{equation}
\label{eq:class2_bloch_sphere_1}
\dfrac{3}{2} \max_{B_3^\epsilon} \expect{B_3^\epsilon}_\psi =  \max_{B_3^\epsilon} \dfrac{\sqrt{3}}{2} \vec{b} \cdot (1,1,1).
\end{equation}

From the argument that the observables move symmetrically towards $\psi$, it is clear that the azimuthal angle of $\vec{b}$ can be set to $\varphi = \pi/4$. In addition, from the fidelity constraint $\tr(B_y^\text{targ}B_y^\epsilon)\geq 2-4\epsilon$, we get that the polar angle $\theta$ fulfills the inequality constraint $\dfrac{1}{2}(1+\cos\theta) \geq 1-\epsilon$. In particular, the optimal choice of $\theta$ corresponds to the case of equality. As a result, the $f(\epsilon)$ is maximised when $B_3^\epsilon$'s Bloch vector is
\begin{equation}
\vec{b} = (\sqrt{2}\sqrt{\epsilon(1-\epsilon)}, \sqrt{2}\sqrt{\epsilon(1-\epsilon)}, 1-2\epsilon),
\end{equation}
which inserted into ~\eqref{eq:class2_bloch_sphere_1} yields
\begin{equation}
\label{eq:class2_bloch_sphere_2}
f(\epsilon) = \dfrac{\sqrt{3}}{2}\left(1+2\sqrt{2\epsilon(1-\epsilon)}-2\epsilon\right)
\end{equation}

We observe that $f(\epsilon)$ increases monotonically from $\sqrt{3}/2$ to $3/2$ as a function of $\epsilon$, where the upper bound corresponds to the local hidden variable bound for the witness. Specifically, from Eq.~\eqref{eq:class2_bloch_sphere_2} we find that the critical magnitude of imprecision at which $f(\epsilon)$ overtakes the standard LHS bound $\beta_0 = 1$ is given by
\begin{equation}
\epsilon^* = \dfrac{9 -2\sqrt{3} - \sqrt{30}}{18} \approx 0.00326.
\end{equation}
 That is, $\beta_0 = 1$ corresponds to the LHS bound until we reach an imprecision $\epsilon^* \approx 0.00326$. Thus, we interpret $\epsilon^*$ as the length of the imprecision plateau.

\subsection{Proof via method of Ref.~\cite{Tavakoli2024}}
We will now show an alternative derivation of $f(\epsilon)$. The main theorem of Ref.~\cite{Tavakoli2024} provides a simple way to determine upper bounds on steering witnesses when subject to imprecise measurements. However, applying that theorem directly to the ESI witness fails to reveal the existence of the plateau because the upper bound on the LHS-value is not tight. Here, we show that the method of Ref.~\cite{Tavakoli2024} can be strengthened to identify the imprecision plateau. The main idea is to apply ``Lemma 1'' in Ref.~\cite{Tavakoli2024} separately to the witness expression for each of the deterministic input-output strategies of Alice.

The Lemma states that for any imprecise measurement $B_{b \mid y}$ at least $\epsilon_{by}$-close in fidelity to the rank-one target measurement $B_{b \mid y}^{\text{targ}}$ it holds that
\begin{equation}
\label{eq:sigma_preceq}
B_{b \mid y} \preceq(1+\mu_{by})B_{b \mid y}^{\text{targ}}+\frac{1}{2}\left(\sqrt{\mu_{by}^2+4 \epsilon_{by} (1+\mu_{by})}-\mu_{by}\right) \mathds{1}
\end{equation}
where $\epsilon_{by}$ is the imprecision parameter associated with the $y$'th measurement of Bob with outcome $b$ and the parameter $\mu_{by} \geq -1$  \cite{Tavakoli2024}. Since Bob's imprecision parameters are assumed to be equal, i.e., $\epsilon_{by} = \epsilon$, this implies that $\mu_{by} = \mu$. Now, applying Eq.~\eqref{eq:sigma_preceq} to $\max_{\psi, B_y^\epsilon}  f^*(\epsilon)$, we find that
\begin{equation}
\label{eq:class2_theorem}
\begin{aligned}
f^*(\epsilon) \leq \min_{\mu \geq -1} \dfrac{\sqrt{3}}{2}(1+\mu)+ \dfrac{3}{2}\sqrt{\mu^2+4\epsilon(1+\mu)} = \tilde{f}(\epsilon)
\end{aligned}
\end{equation}
where the state was optimally selected as $\psi = \big(\mathds{1} + (1,1,1)/\sqrt{3}\cdot \vec{\sigma}\big)/2$. Next, we solve $\dfrac{\partial \tilde{f}(\epsilon) }{\partial \mu} = 0$ to find the minimum. Of the two possible solutions, we observe that the one corresponding to the smaller value of $ \tilde{f}(\epsilon)$ is 
\begin{equation}
\label{eq:class2_theorem_2}
\mu_{opt} = -2\epsilon-\sqrt{2}\sqrt{\epsilon(1-\epsilon)}.
\end{equation}
Evaluating $ \tilde{f}(\epsilon)$ at $\mu = \mu_{opt}$ yields
\begin{equation}
\dfrac{\sqrt{3}}{2}\left(1+2\sqrt{2\epsilon(1-\epsilon)}-2\epsilon\right),
\end{equation}
and we thereby recover the result in Eq.~\eqref{eq:class2_bloch_sphere_2}.

\section{Family of plateau witnesses}\label{AppCon}
We derive a family of steering witnesses that feature an imprecision plateau. The family is parameterised by an integer $n\geq 3$. Alice is given $n_X=2^{n-1}$ inputs and Bob is given $n$ inputs. The outcomes are binary. We select Bob's target observables to be pairwise anti-commuting, i.e.~$\{B^\text{targ}_j,B^\text{targ}_k\}=0$ for all $j\neq k$. We choose to represent Alice's input $x$ as a string  $x=(0,x_2,...,x_{n})$ where $x_i \in\{0,1\}$ for $i=2,\ldots,n$. A generic full-correlation witness takes the form
\begin{equation}\label{eq:Witness_family}
\mathcal{W}_0=\sum_{x=1}^{2^{n-1}}\sum_{y=1}^n c_{xy}\expect{A_x\otimes B^\text{targ}_y}.
\end{equation}
We will select 
\begin{equation}
c_{xy} = \frac{(-1)^{x_y}}{n_X}.
\end{equation}
Note that this choice ensures that the first column of the coefficient matrix sums to one, whereas all other columns individually sum to zero. 

Below, we first show a quantum strategy for this witness, then we prove the LHS bound  and finally we prove that the witness features an imprecision plateau for every $n$.

\subsection{Quantum strategy}\label{Qstrat}
Let us first consider a quantum model for $\mathcal{W}_0$. We select Alice observables as $A_x = \frac{1}{\sqrt{n}}\sum_{j=1}^n (-1)^{x_j}  (B_j^\text{targ})^T$ and the shared state as maximally entangled ($\phi^+$) \cite{Chailloux2016}. Then the witness becomes
\begin{equation}
\begin{aligned}
\mathcal{W}_0^{Q} &= \dfrac{1}{n_X \sqrt{n}} \sum_{y, \bar{y}=1}^n \left(\langle (B^{\text{targ}}_{\bar{y}})^T \otimes B^{\text{targ}}_y \rangle_{\phi^+} \sum_x (-1)^{x_{\bar{y}} + x_y}\right).
\end{aligned}
\end{equation}
We then use that $\sum_x (-1)^{x_{\bar{y}} + x_y}  = n_X \delta_{y,\bar{y}}$. This gives
\begin{align}
	\mathcal{W}_0^{Q} &= \dfrac{1}{\sqrt{n}} \sum_{y=1}^n \langle (B^{\text{targ}}_{y})^T \otimes B^{\text{targ}}_y \rangle_{\phi^+} \\
	& =\frac{1}{d\sqrt{n}}\sum_y\tr(\openone_d)=\sqrt{n},
\end{align}
where in the second step we used the property of the maximally entangled state that $\expect{O_1\otimes O_2}_{\phi^+}=\frac{1}{d}\tr(O_2O_1^T)$ where $d$ is the local dimension of $\phi^+$, and in the third step we used the property of observables; that $(B^\text{targ}_y)^2=\openone$.

\subsection{LHS bound}
Consider now an arbitrary deterministic strategy for Alice to decide her output, namely $f_\lambda(x)\in\{0,1\}$ where $\lambda$ indexes the choice of function. For given $\lambda$, the corresponding optimal value of $\mathcal{W}_0$ becomes. 
\begin{align}
\mathcal{W}_0^\lambda&=\nonumber
\frac{1}{n_X}\|\sum_{x,y} (-1)^{f_\lambda(x)+x_y} B^\text{targ}_y\|_\infty=\| \sum_y t_y^\lambda B^\text{targ}_y \|_\infty\\\nonumber
&= \sqrt{\| \big(\sum_y t_y^\lambda B^\text{targ}_y\big)^2 \|_\infty}=\sqrt{\| \sum_y (t_y^\lambda)^2 \openone \|_\infty}\\
& =\sqrt{\sum_y (t_y^\lambda)^2}=\|\vec{t}_\lambda\|_2.
\end{align}
where we defined $t_y^\lambda=\frac{1}{n_X}\sum_x (-1)^{f_\lambda(x)+x_y}$ and $\vec{t}_\lambda=(t_1^\lambda,\ldots,t_n^\lambda)$. In the above we have used the anti-commutativity of the target observables. The LHS bound for the steering witness then becomes
\begin{equation}\label{eq:t_vec}
\mathcal{W}_0\stackrel{\text{LHS}}{\leq} \max_\lambda \|\vec{t}_\lambda\|_2
\end{equation}
We can now compute the vector norms as
\begin{equation}
\begin{aligned}
 \|\vec{t}_\lambda\|^2_2 &= \sum_y \bigg(\frac{1}{n_X}\sum_x (-1)^{f_\lambda(x)+x_y}\bigg)^2\\
&=\dfrac{1}{n_X^2} \sum_{x,z} (-1)^{f_\lambda(x)+f_\lambda(z)}\sum_y  (-1)^{x_y+z_y}.
\end{aligned}
\end{equation}
Define an $n_X \times n_X$ matrix whose entries are $S_{x,z}=\sum_y (-1)^{x_y+z_y}$. We also define normalised vectors $\ket{u_\lambda}=\frac{1}{\sqrt{n_X}}\sum_x (-1)^{f_\lambda(x)}\ket{x}$. Then we can write
\begin{equation}
\|\vec{t}_\lambda\|^2_2= \frac{1}{n_X} \bracket{u_\lambda}{S}{u_\lambda}.
\end{equation}
The vectors $\ket{u_\lambda}$ are constrained to have all their entries equal to $\pm1/\sqrt{n_X}$. To avoid having to deal with each $\lambda$, we relax this structure to $\ket{u}$ being a generic normalised vector. This entails relaxing the above into a largest-eigenvalue expression
\begin{equation}
\max_\lambda \|\vec{t}_\lambda\|^2_2 \leq  \frac{1}{n_X} \|S\|_\infty.
\end{equation}
To determine the largest eigenvalue of $S$, we make the observation that $S$ is an unnormalised projector, i.e.~that it can be written as $S=n_X \Pi$ where $\Pi$ is a projector. This directly leads to 
$\|S\|_\infty=n_X$ and therefore to the LHS bound 
\begin{equation} \label{eq:bound_lhs}
\mathcal{W}_0\stackrel{\text{LHS}}{\leq} 1.
\end{equation}
We are left with proving that $\Pi=S/n_X$ is a projector, meaning that it fulfills the relation $\Pi^2 = \Pi$. Indeed we find that,
\begin{equation}
\begin{aligned}
\Pi^2 &= \dfrac{1}{n_X^2} \sum_{x,z} S_{x,z} |x\rangle \langle z | \sum_{x',z'} S_{x',z'} |x' \rangle \langle z' |\\
&= \dfrac{1}{n_X^2} \sum_{x,z'} \sum_{y,y'}  (-1)^{x_y+ z'_{y'}} \underbrace{\bigg( \sum_z (-1)^{z_{y'} + z_y} \bigg)}_{=n_X \delta_{y,y'}} |x\rangle  \langle z' | \\
&= \dfrac{1}{n_X} \sum_{x,z'} \underbrace{\bigg( \sum_y(-1)^{x_y+ z'_y} \bigg)}_{=S_{x,z'}} |x\rangle  \langle z' | = \frac{S}{n_X}=\Pi.
\end{aligned}
\end{equation}

\subsection{Existence of imprecision plateau}
We show that the steering witness $\mathcal{W}_0$ supports an imprecision plateau for every $n$. Start by considering the specific set of deterministic strategies of the form $\{f_\lambda(x) = x_{\bar{y}}\}$, where $\bar{y} \in [n]$. This class of strategies results in 
\begin{equation}\label{eq:wit_opt}
\mathcal{W}_0 ^{f_\lambda(x) = x_{\bar{y}}} = \dfrac{1}{n_X} \sum_{x,y}  (-1)^{x_{\bar{y}}+x_y} \langle B_{y}^\text{targ} \rangle_{\psi} = \langle B_{\bar{y}}^\text{targ} \rangle_{\psi},
\end{equation}
\noindent where we again have used that $\sum_x (-1)^{x_{\bar{y}}+x_y}  = n_X \delta_{y,\bar{y}}$. By optimising the witness with respect to Bob's local state $\psi$, it is clear that it saturates the LHS bound
\begin{equation}
\max_\psi \langle B_{\bar{y}}^\text{targ} \rangle_{\psi} = ||\vec{t}||_2 = 1.
\end{equation}

We now want to show that no other class of deterministic strategies can reach the steering bound. Since the LHS bound is tight, Eq.\eqref{eq:t_vec} gives that it is enough to determine which vectors $\vec{t}_\lambda = (t_1^\lambda,..., t_n^\lambda)$ that can reach unit norm.

Assume w.l.g.~that $t_1^\lambda$ is the largest element in $\vec{t}_\lambda$ and denote its value by $t_1^\lambda  = (n_X-2\alpha)/n_X$. Note that $t_1^\lambda=1$ implies $t_i^\lambda=0$ for all $i\neq 1$. The integer parameter $\alpha \in [0,n_X/4]$ is the number of positions at which the output  $(-1)^{f_\lambda(x)}$ does not match the value of the first column of the coefficient matrix $c_{x,y=1} = (1,...,1)^T$. By extension, this means that $(-1)^{f_\lambda(x)}$ can match the second column $c_{x,y=2}$ up to at most $\alpha$ additional positions, increasing the value of $t_2^\lambda$ from zero to $2\alpha$. The same holds for the remaining columns. However, the lexicographic structure of the $n-1$ last columns in $c_{xy}$ impose additional constraints on $t^\lambda_y$. By inspection of $c$, we find that the optimal combination of column sums for maximising the norm $\|\vec{t}_\lambda\|_2$ cannot be better than 
\begin{equation}\label{eq:t}
\vec{t}_\lambda \leq \dfrac{1}{n_X}(n_X-2\alpha, 2\alpha, 2\alpha, \min \{2\alpha,2^{n-3}\},...,\min \{2\alpha,2\}).
\end{equation}
\noindent Here the elements have been sorted in descending order. From \eqref{eq:t} we can compute an upper bound of the vector norm, which yields
\begin{align}
\|\vec{t}_\lambda\|_2 \leq \dfrac{1}{n_X}\bigg( (n_X-2\alpha)^2 + 2(2\alpha)^2 + \sum_{y=4}^{n}  (\min \{2\alpha,2^{n+1-y}\})^2 \bigg)^{\frac{1}{2}}
\end{align}
In the interval $\alpha\in[0,n_X/4]$ this function has a unique minimum obtained from finding the roots of the $\alpha$-derivative. Thus the maximum is obtained at one of the the end-points, namely either $\alpha=0$ or $\alpha=n_X/4$. The former gives $\|\vec{t}_\lambda\|=1$. This is expected since $\alpha = 0$ corresponds to the case in which the deterministic strategies take the form $\{f_\lambda(x) = x_{\bar{y}}\}$. In contrast, at the other end of the interval, where $\alpha = n_X/4$, the vector norm is bounded by
\begin{equation}
\begin{aligned}
&\|\vec{t}_\lambda\|_2 \leq \bigg(3(\dfrac{1}{2})^2 + \sum_{y=4}^{n} \dfrac{(2^{n+1-y})^2}{n_X^2} \bigg)^{1/2} = \sqrt{\dfrac{5}{6}-\dfrac{4}{3n_X^2} }<  1.
\end{aligned}
\end{equation}

That is, by studying $\vec{t}_\lambda$, we find that the optimal deterministic strategies are on the form $f_\lambda(x) = x_{\bar{y}}$. Importantly, these strategies uniquely attain the LHS bound of the steering witness $\mathcal{W}_0$ and they imply no $\epsilon$-dependence once the measurements become $\epsilon$-imprecise because $\mathcal{W}_0^\lambda$ depends only on a single observable of Bob. Since this holds for all $n \geq 3$, where the special case $n=3$ is the ESI witness, it implies that an imprecision plateau exists for every $n$.

\subsection{Example: the case of $n=4$}\label{AppDim4}

We consider now in further detail the simplest case beyond the ESI, namely that corresponding to $n=4$. This leads to a steering inequality supporting an imprecision plateau that is optimally violated using four-dimensional maximal entanglement.

We associate Bob's four inputs, $y=1,2,3,4$, with anti-commuting target observables $B^\text{targ}_1=X\otimes X$, $B^\text{targ}_2=Y\otimes X$, $B^\text{targ}_3=Z\otimes X$, $B^\text{targ}_4=\openone\otimes Y$. The steering inequality then becomes 
\begin{equation}\label{witness}
\mathcal{W}_0= \sum_{x_2,x_3,x_4,y}(-1)^{x_y}\expect{A_x\otimes B_y}\stackrel{\text{LHS}}{\leq} 1.
\end{equation}
By distributing a four-dimensional maximally entangled state and selecting suitable measurements for Alice (see section~\ref{Qstrat}), one can achieve the violation $\mathcal{W}^Q_0=2$. It is also interesting to note that local hidden variable models cannot exceed $\mathcal{W}\leq 3/2$. 

Since Alice has eight inputs and binary outputs, she has 256 deterministic input-output strategies. We have evaluated $\mathcal{W}_0$ for each of these and found that they reduce into just seven qualitatively different classes (up to permutations of inputs, outputs and symmetry consideration). This is in analogy with the three classes appearing for the ESI in the main text. The seven classes are as follows.
\begin{equation}\label{classes_four}
\begin{aligned}
\mathcal{W}_\epsilon &= \langle B_1 \rangle_\psi &&=g^*(\epsilon) \\
\mathcal{W}_\epsilon &= \langle 3B_1 + B_2 + B_3 + B_4\rangle_\psi/4 \\
\mathcal{W}_\epsilon &= \langle B_1+B_2+B_3 \rangle_\psi/2 &&= f^*(\epsilon) \\
\mathcal{W}_\epsilon &= \langle B_1+B_2 \rangle_\psi/2\\
\mathcal{W}_\epsilon &= \langle B_1+B_2+B_3 + B_4 \rangle_\psi/4\\
\mathcal{W}_\epsilon &= \langle B_1 \rangle_\psi/2\\
\mathcal{W}_\epsilon &= 0
\end{aligned}
\end{equation}

Three of these classes have no $\epsilon$-dependence, in particular the first class $g(\epsilon)\equiv \max g^*(\epsilon) = 1$, which gives the standard LHS bound in \eqref{witness}. The remaining four classes are $\epsilon$-dependent. Among these four, the second and third strategies have the same LHS value and they exceed that of the other two strategies. But the third strategy grows faster in $\epsilon$ than does  second and therefore we must consider $f^*(\epsilon)$.

We compute the plateau length using the fidelity quantifier of imprecision. To this end we evaluate $f(\epsilon)=\max f^*(\epsilon)$. Since we no longer can rely on Bloch sphere symmetries, we rely instead on our generalisation of the Lemma introduced in Ref.~\cite{Tavakoli2024}. Our generalisation applies to target measurements of higher rank than one. This is discussed in Appendix.~\ref{AppGen}. Here, we directly use the result, which is that
\begin{equation}
\label{eq:theo_app4}
B_{b\mid y} \preceq(1+\mu) B_{b\mid y}^{\text{targ}}+\frac{\sqrt{\mu^2+4 (2\epsilon)(1+\mu)}-\mu}{2} \mathds{1},
\end{equation}
for all $ \mu \geq -1$. However, since Bob's target measurements are products of single-qubit observables, we consider the lab measurement to factor by each single-qubit observable having an $\tilde{\epsilon}$-imprecision (except the identity observable). This implies that $\epsilon = 2\tilde{\epsilon}$. Inserting \eqref{eq:theo_app4} into the expression for $f^*(\epsilon)$, we find that the expectation value over the target observables $B_y^\text{targ} \in \mathcal{H}_1\otimes \mathcal{H}_2$ can be decomposed as follows
\begin{equation}
\langle B_1^\text{targ}+B_2^\text{targ}+B_3^\text{targ}\rangle_\psi = \operatorname{Tr}\big((X+Y+Z)\operatorname{Tr}_{\mathcal{H}_2}(\mathds{1} \otimes X \psi)\big).
\end{equation}

We recognize that $\psi$ is optimally chosen such that partial trace over the second subsystem becomes
\begin{equation}
\operatorname{Tr}_{\mathcal{H}_2}(\mathds{1} \otimes X \psi) = \dfrac{\mathds{1}+(1,1,1)/\sqrt{3} \cdot \vec{\sigma}}{2}.
\end{equation}

At this point, the task to evaluate the upper bound of $f^*(\epsilon)$ is analogous to the procedure described in section~\ref{AppClass2}. Following this, we find that
\begin{equation} \label{eq:inc_app4}
f(\epsilon) = \max_{\psi, B_y^{\epsilon_y}} f^*(\vec{\epsilon})  \leq \dfrac{\sqrt{3}}{2}+\sqrt{6}\sqrt{2\epsilon(1-2\epsilon)}-2\sqrt{3}\epsilon = \tilde{f}(\epsilon),
\end{equation}
In fact, for every value of $\epsilon$, we find that $\tilde{f}(\epsilon)$ matches the value of $f(\epsilon)$. This can be verified by computing $\tilde{f}(\epsilon)$ through the see-saw method. This implies that $\tilde{f}(\epsilon)$ gives a tight bound. Lastly, solving $\tilde{f}(\epsilon) = 1$, we find that the threshold, at which the third class overtakes the first, is given by
\begin{equation}
\tilde{\epsilon}^* = \dfrac{9-2\sqrt{3}-\sqrt{3}}{32}.
\end{equation}
Specifically, the threshold error of each qubit is then $\tilde{\epsilon}^*\approx 8.1\times 10^{-4}$. This is large enough to encompass the imprecisions for $X$ and $Z$ Pauli operators experimentally observed in Ref.~\cite{Cao2023}.

\section{Approximating the plateau length of the ESI witness for \texorpdfstring{$\vec{\epsilon} = (\epsilon_X,\epsilon_Y,\epsilon_Z)$}{TEXT}}\label{AppTaylor}
Consider the ESI scenario in which each of Bob's three measurements is associated with an independent imprecision parameter, such that $\vec{\epsilon} = (\epsilon_X,\epsilon_Y,\epsilon_Z)$. By optimizing $f(\vec{\epsilon}) = \max_{\psi, B_y^{\epsilon_y}} f^*(\vec{\epsilon})$ we can now derive a plateau region by solving $f(\epsilon) = 1$ for one of the imprecision parameters, say $\epsilon_Z$, in terms of the other two $\epsilon_X, \epsilon_Y$. Although we cannot find an exact analytical solution of this problem we can still approximate it using other methods. We discuss two of them below, but neither is exact. This motivates our later use of SDP methods, which turn out to easily provide more accurate results.

\subsection{Approximation via method of Ref.~\cite{Tavakoli2024}}
To calculate an upper bound of the value of $f(\vec{\epsilon})$  we start by using the theorem from Ref.~\cite{Tavakoli2024}. Following the same procedure as in section~\ref{AppClass2}, applying ~\eqref{eq:sigma_preceq} to each of Bob's observables in $f^*(\vec{\epsilon})$, with $\epsilon_{by} = \epsilon_y$, yields
\begin{equation}\label{eq:formula}
\begin{aligned}
f^*(\vec{\epsilon})\leq &\max_{\psi} \min_{\mu_i \geq -1} \dfrac{1}{2} \bigg\langle \sum_{i=1}^3 (1+\mu_i)B_i^{\text{targ}} \bigg\rangle_{\psi} \\
&+ \dfrac{1}{2} \sum_{i=1}^3 \sqrt{\mu_i^2+4\epsilon_i(1+\mu_i)}.
\end{aligned}
\end{equation}
Then, the optimal state $\psi$ is the one that aligns with $ \sum_{i=1}^3 (1+\mu_i)B_i^{\text{targ}} = (1+\mu_1,1+\mu_2,1+\mu_3)\cdot \vec{\sigma}$. This gives that
\begin{equation}\label{eq:upp_bound_eps}
\begin{aligned}
f^*(\vec{\epsilon})& \leq \min_{\mu_i \geq -1} \dfrac{1}{2} \sqrt{\sum_{i=1}^3 (1+\mu_i)^2} \\
&+ \dfrac{1}{2} \sum_{i=1}^3 \sqrt{\mu_i^2+4\epsilon_i(1+\mu_i)} = \tilde{f}(\epsilon).
\end{aligned}
\end{equation}

At this point it is challenging to analytically constrain the upper bound further. Thus, to compute the plateau region by solving  $\tilde{f}(\epsilon) = 1$, we numerically optimize \eqref{eq:upp_bound_eps} over $\mu_i$. That is, for fixed values of $(\epsilon_X, \epsilon_Y)$, we can find the length of the imprecision plateau in terms of $\epsilon_Z$ by successively increasing the it until $\tilde{f}(\epsilon)$ reaches one. This gives a plateau region $\epsilon^*(\epsilon_X,\epsilon_Y)$. Although the bounds found are not tight, they are strictly upper bounds of the associated LHS bounds. In the range $\epsilon_X, \epsilon_Y \in [0,0.01]$ the estimated plateau length obtained via the formula \eqref{eq:upp_bound_eps} at worst deviates from the values obtained by brute-force optimization by a factor $O(10^{-3})$.  Considering that the plateau length is of the same order, $O(10^{-3})$, this deviation is quite significant. At best, the deviation is of the order $O(10^{-5})$. 

\subsection{Taylor expansion}
Alternatively, the plateau region can be approximated using Taylor expansion.   Associating each of Bob's lab measurement $B_i$ with the Bloch vector $\vec{b}_i = (b_{ix},b_{iy},b_{iz})$, we observe that
\begin{equation} \label{eq:EBI_inc_general}
f(\vec{\epsilon}) = \max_{\psi, B_y^{\epsilon_y}} \dfrac{1}{2}\langle B_1+B_2+B_3\rangle_{\psi} =\max_{\psi, B_y^{\epsilon_y}} \dfrac{1}{2} \vec{\textbf{b}} \cdot\vec{\psi} = \max_{B_y^{\epsilon_y}}  \dfrac{1}{2} |\vec{\textbf{b}}|
\end{equation}
 where $ \vec{\textbf{b}} =  (\sum_i b_{ix}, \sum_i b_{iy}, \sum_i b_{iz})$. Moreover, Eq.~\eqref{eq:EBI_inc_general} is maximised when the deviation between the lab and the associated target measurement is taken to be as large as possible. Given this, the fidelity constraint then yields that the Bloch vectors $\vec{b_i} = (\sin{\theta_i}\cos{\varphi_i}, \sin{\theta_i}\sin{\varphi_i}, \cos{\theta_i})$ optimally are chosen as
\begin{align*}
\vec{b}_x &= (a_x, \sqrt{\sin{\theta_x}^2-a_x^2}, \cos{\theta_x})\\
\vec{b}_y &= (\sqrt{\sin{\theta_y}^2-a_y^2},a_y,\cos{\theta_y})\\
\vec{b}_z &= (\sqrt{1-a_z^2}\cos{\varphi_z}, \sqrt{1-a_z^2}\sin{\varphi_z}, a_z)
\end{align*}
where we have denoted $a_i = 1-2\epsilon_i$ for $i= x,y,z$. Moreover, we require that $\sin{\theta_j} \geq a_j$ for $j =x,y$. Next, inserting the the optimal $\vec{b}_i$ into the Eq.~\eqref{eq:EBI_inc_general}, we find that
\begin{equation}
\begin{aligned}
\label{eq:EBI_inc_general_2}
f(\vec{\epsilon}) = \max_{\theta_x,\theta_y,\varphi_z} 2 &\bigg((a_x+\sqrt{\sin{\theta_y}^2-a_y^2}+\sqrt{1-a_z^2}\cos{\varphi_z})^2 \\
+ &(\sqrt{\sin{\theta_x}^2-a_x^2} + a_y + \sqrt{1-a_z^2}\sin{\varphi_z})^2 \\
+ &(\cos{\theta_x}+\cos{\theta_y}+a_z)^2\bigg)^{1/2}.
\end{aligned}
\end{equation}
At this point, we want to compute the plateau region in terms of $\epsilon_Z$ for fixed pair of $(\epsilon_X, \epsilon_Y)$. Thus, solving $ f(\vec{\epsilon})= 1$ in terms of $\epsilon_Z$ yields
\begin{equation}
\label{eq:az}
\epsilon_Z(\epsilon_X,\epsilon_Y) = \min_{\theta_x,\theta_y,\varphi_z} \dfrac{1}{2}\bigg(1-\dfrac{d_2+\sqrt{d_2-d_1d_3}}{2d_1}\bigg)
\end{equation}
 where we have denoted
\begin{equation}
\begin{aligned}
\label{eq:notation_diff_eps_taylor}
d_1 &= (\cos{\theta_x}+\cos{\theta_y})^2+e_1^2 \\
d_2 &= (\cos{\theta_x}+\cos{\theta_y})e_2\\
d_3 &= e_2^2-4e_1^2\\
e_1 &= a_x\cos{\varphi_z}+a_y\sin{\varphi_z}+\sqrt{\sin{\theta_y}^2-a_y^2}\cos\varphi_z\\
&+\sqrt{\sin{\theta_x}^2-a_x^2}\sin{\varphi_z}\\
e_2 &= 1-2(a_x \sqrt{\sin{\theta_y}^2-a_y^2} + a_y\sqrt{\sin{\theta_x}^2-a_x^2}\\
& + \cos{\theta_x}\cos{\theta_y}).
\end{aligned}
\end{equation}

Under the assumption that the measurement imprecisions are small, we can now employ Taylor expansion to approximate the trigonometric functions. In accordance with the target observables, we choose the expansion points to be $\theta_x: \pi/2, \theta_y: \pi/2$ and $\varphi_z: \pi/4$. From these, Taylor expansion up to first order gives that
\begin{equation}
\label{eq:expansion_points}
\begin{aligned}
\sin{\theta_x} &\approx 1-\epsilon_X\\
\sin{\theta_y} &\approx 1-\epsilon_Y\\ 
\sin{\varphi_z} &\approx \dfrac{1+\sqrt{\epsilon_X/2}-\sqrt{\epsilon_Y/2}}{\sqrt{2}} = s_z
\end{aligned}
\end{equation}
 Inserting the approximations \eqref{eq:expansion_points} back into the variables $e_1$ and $e_2$ we obtain that
\begin{equation} 
\label{eq:approx_ei}
\begin{aligned}
e_1 &\approx (1-2\epsilon_X) c_z +(1-2\epsilon_Y)s_z+\sqrt{\epsilon_Y(2-3\epsilon_Y)}c_z \\
&+\sqrt{\epsilon_X(2-3\epsilon_X)}s_z\\
e_2 &\approx 1-2((1-2\epsilon_X) \sqrt{\epsilon_Y(2-3\epsilon_Y)}\\
&+ (1-2\epsilon_Y)\sqrt{\epsilon_X(2-3\epsilon_X)} \\
&+ \sqrt{\epsilon_X(2-\epsilon_x)}\sqrt{\epsilon_Y(2-\epsilon_Y)})
\end{aligned}
\end{equation}

\noindent where we have denoted $c_z = \sqrt{1-s_z^2}$. We note that the final expressions only depends on the imprecision parameters $\epsilon_X$ and $\epsilon_Y$. Thus, using Taylor expansion we eliminate the plateau length's dependence on the original optimization parameters $(\theta_x,\theta_y,\varphi_z)$. At worst, this methods gives that the values of the plateau lengths deviates from the numerical optimized bounds at an order of $O(10^{-4})$ for $\epsilon_X, \epsilon_Y \in [0,0.01]$. However, for most input pairs $(\epsilon_X, \epsilon_Y)$ the values only differ at an order of $O(10^{-5})$. 

\begin{figure}
	\centering
	\includegraphics[width=\columnwidth]{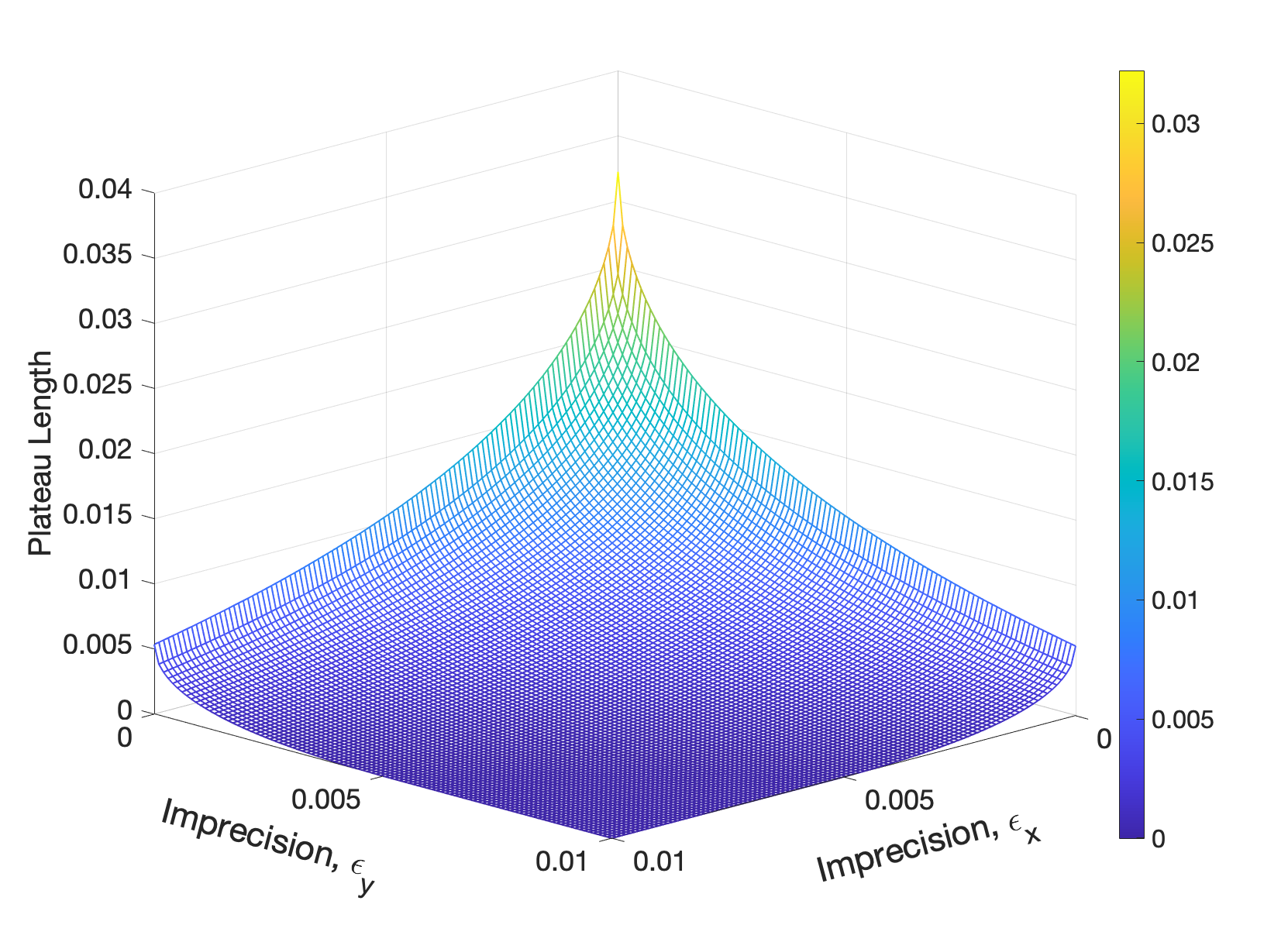}
	\caption{Imprecision plateau for the ESI witness where each of the trusted parties measurements are associated with an independent imprecision parameter. The plateau length in $\epsilon_Z$ on the z-axis is given as a function of $(\epsilon_X,\epsilon_Y)$.}\label{Fig_xyz}
\end{figure}

\section{Semidefinite programming method}\label{AppSDP}

\subsection{Bounding witnesses}
We discuss an SDP relaxation method to bound arbitrary linear witnesses in an LHS model with imprecise measurements. Define an operator list $L=\{\openone, \sigma, \{B_{b|y}^\text{targ}\},\{B_{b|y}^{\vec{\epsilon}}\}\}$. Over $L$, define an arbitrary monomial list, $\mathcal{S}$, with products of at least length one (i.e. the list $L$). Define the  $|\mathcal{S}|\times |\mathcal{S}|$ moment matrix as 
\begin{equation}
\Gamma_{u,v}=\tr(u^\dagger v)
\end{equation}
for $u,v\in\mathcal{S}$. By construction, we have $\Gamma\succeq 0$. In order to impose that all operators are $d$-dimensional, we use the sampling heuristic proposed in \cite{Navascues2015}, and to impose the imprecision conditions $\tr(B_{b\mid y}^\epsilon B_{b\mid y}) \leq 1-\epsilon_{by}$ we use ideas of \cite{Tavakoli2021}. That is, we sample $L$ randomly over $\mathbb{C}^d$, with the exception of $\{B_{b|y}^\text{targ}\}$. This procedure is repeated independently and the associated moment matrix samples are denoted $\{\Gamma^{(i)}\}_{i=1}^n$. The routine is terminated when the next sample is linearly dependent on the set of previously generated samples. The final moment matrix is then an affine combination over this basis, 
\begin{align}
& \Gamma=\sum_{i=1}^n s_i \Gamma^{(i)} \quad \text{with}&&\sum_i s_i=1,
\end{align}
for some coefficients $\{s_i\}$. The imprecision conditions can then be imposed as linear inequalities, namely 
\begin{equation}
\Gamma_{B_{b|y}^\text{targ},B_{b|y}^{\vec{\epsilon}}}\geq 1-\epsilon_{by}.
\end{equation}
Over this domain, we optimise the witness value for LHS models with imprecise measurements. This is done independently for every deterministic strategy of Alice, characterised by the response functions $D_\lambda(a|x)\in\{0,1\}$. Thus, we write
\begin{equation}
W^\lambda_{\vec{\epsilon}}=\sum_{a,b,x,y}c_{abxy}D_\lambda(a|x)\Gamma_{\sigma,B_{b|y}^{\vec{\epsilon}}},
\end{equation}
This constitutes an SDP. Solving the SDP for every deterministic strategy $\lambda$, we obtain an upper bound $\mathcal{W}^\lambda_{\vec{\epsilon}}\leq \alpha^\lambda$. Our final bound on the witness is then obtained from
\begin{equation}
\mathcal{W}_{\vec{\epsilon}}\leq \max_\lambda\{\alpha^\lambda\}.
\end{equation}

In the main text, we discussed applying this method to bound the plateau length for the ESI witness when each of Bob's measurements has an independent imprecision parameter, $\vec{\epsilon}=(\epsilon_X,\epsilon_Y,\epsilon_Z)$. In Fig.~\ref{Fig_xyz}, we illustrate the plateau length in $\epsilon_Z$ as a function of $(\epsilon_X,\epsilon_Y)$. All values match those obtained with seesaw search (interior point method) to five decimals. Therefore, this can for practical purposes be seen as a proof of the plateau length.

\subsection{Imprecise LHS models for correlations}
The SDP relaxations can be further extended to apply directly to probability distributions, $p_\text{targ}(a,b|x,y)$, without any reference to a specific witness. For this purpose, we associate a set of variables $\{s_i^\lambda\}_i$ to each determinstic strategy $\lambda$, and a corresponding moment matrix 
\begin{align}
&\Gamma^\lambda =\sum_{i=1}^n s_i^\lambda \Gamma^{(i)} \quad\text{with}&& \sum_{i,\lambda}s_i^\lambda=1,
\end{align}
where $\Gamma^{(i)}$ still is the sampled moment matrix basis. We require that $\Gamma^\lambda\succeq 0$.  The imprecision condition takes the form 
\begin{equation}
\Gamma^\lambda_{B_{b|y},B^{\vec{\epsilon}}_{b|y}}\geq (1-\epsilon_{by})q(\lambda),
\end{equation}
where $q(\lambda)=\sum_i s_i^\lambda$ is the probability of running strategy $\lambda$.

As a natural benchmark, we use an isotropic noise model, corresponding to the constraints $\forall a,b,x,y$, 
\begin{equation}
vp_\text{targ}(a,b|x,y)+\frac{1-v}{N^2}=\sum_{a,b,x,y}\sum_\lambda c_{abxy}D_\lambda(a|x)\Gamma^{\lambda}_{\sigma,B^{\vec{\epsilon}}_{b|y}},
\end{equation}
where $N$ is the number of outputs per party. Over this SDP domain, we optimise $v\in[0,1]$. In order to prove that a given set of imprecision parameters $\vec{\epsilon}$ falls within a plateau, we must find that the solution of the SDP is identical to the critical visibility for standard steering with idealised measurements. The latter can be easily computed \cite{Cavalcanti2017}.

To test it, we have considered $p_\text{targ}$ obtained from Alice and Bob sharing $\ket{\phi^+}$ and both measuring the three directions corresponding to an equilateral triangle in the XZ-plane of the Bloch sphere. The Bloch vectors are $(1,0,0)$, $(-\frac{1}{2},0,\frac{\sqrt{3}}{2})$ and $(-\frac{1}{2},0,-\frac{\sqrt{3}}{2})$ for both parties.  When $\epsilon=0$, these correlations have a steering visibility of $v=\frac{2}{3}$. Using the first level SDP relaxation, we find an imprecision plateau with length $\epsilon^*\approx 0.0287$. This is nearly an order of magnitude larger than the plateau obtained for the ESI witness, but notably its noise tolerance for detecting $\rho_v$ is also significantly worse.

\section{Detection efficiency}\label{AppEta}
Consider steering with imprecise measurements with lossy detection. Alice's measurements succeed only with probability $\eta$. With probability $1-\eta$ she records a failed detection. She can map this by classical post-processing back into one of her outputs. The goal is then to select a quantum strategy that permits detection of steerability for the smallest possible $\eta$ and realistic values of imprecision $\epsilon$. Below, we consider this question first in for the ESI and Pauli witnesses and then for the plateau witness outlined in section~\ref{AppDim4} corresponding to $n=4$.

\subsection{ESI and Pauli witnesses}
\begin{figure}[t]
	\centering
	\includegraphics[width=0.95\columnwidth]{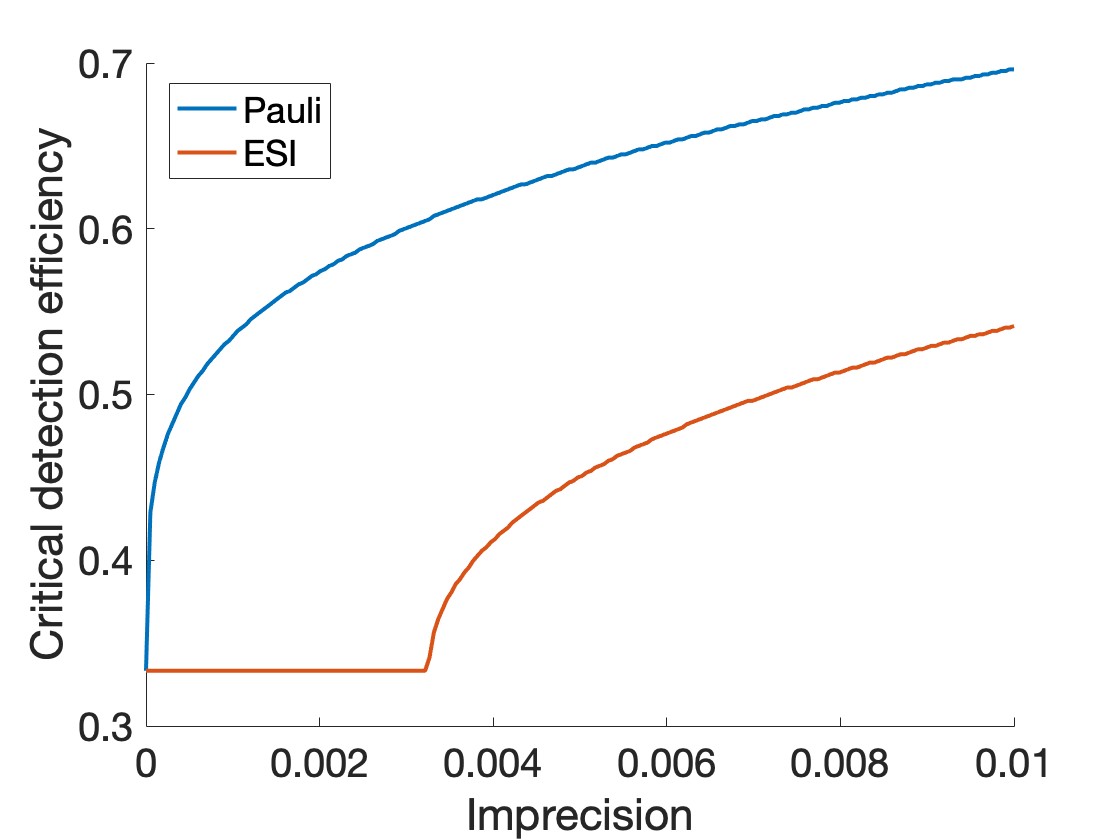}
	\caption{We illustrate the critical detection efficiency $\eta_{\text{crit}}$ for the Pauli and the ESI witness as a function of $\epsilon$.}\label{Fig_deteff}
\end{figure}

Consider the ESI and Pauli witness with lossy detection on Alice's side. We want to compute the critical detection efficiency $\eta_{\text{crit}}$, required for Bob to detect steering. For the ESI witness, let the shared state be the partially entangled state $\ket{\psi_\theta}=\cos\theta\ket{00}+\sin\theta\ket{11}$. Moreover, select Alice's four measurements as corresponding to the Bloch vectors
\begin{equation}
\begin{aligned}
\vec{a}_1 &= (t,t,1)/\sqrt{t^2+1}, &\vec{a}_2 &= (t,-t,-1)/\sqrt{t^2+1},\\
\vec{a}_3 &= (-t,t,-1)/\sqrt{t^2+1}, &\vec{a}_4  &=(-t,-t,1)/\sqrt{t^2+1},
\end{aligned}
\end{equation}
 where the parameter. $t=\sin(2\theta)$. This gives the witness value 
\begin{equation}
\mathcal{W}_0=\sqrt{2-\cos(4\theta)}.
\end{equation}
When her measurement fails, Alice outputs determinstically according to the strategy $ \vec{a}_x = (a_1,a_2,a_3,a_4)\rightarrow (0,1,1,0)$ which yields the witness value $C_\theta=2\cos^2\theta -1$. Solving the equation 
\begin{equation}
\eta \mathcal{W}_0+(1-\eta)C_\theta=1,
\end{equation}
where we have the standard LHS bound on the right hand side, yields that the critical detection efficiency is given by
\begin{equation}
\eta_\text{crit}= \dfrac{2\sin^2\theta}{\sqrt{2-\cos(4\theta)}-\cos(2\theta)}
\end{equation}
valid for $\theta\in[0,\frac{\pi}{4}]$. In the limit $\theta\rightarrow 0$, we obtain the optimal value $\eta_\text{crit}=1/3$. In comparison, the same critical detection efficiency can be achieved with the Pauli steering witness. However, in the presence of imprecise measurements, the critical detection efficiency of the former stays constant within the plateau length whereas the latter rapidly deteriorates, see Figure \ref{Fig_deteff}.

Moreover, the advantage owing to the plateau persists even when comparing ESI to steering inequalities with more settings, which for perfect measurements can arbitrarily reduce $\eta_\text{crit}$. For example, consider the ten-setting steering witness $\mathcal{W}_0=\frac{1}{10}\sum_{x=1}^{10} \expect{A_x B_x^\text{targ}}$, where Bob's target measurements correspond to the vertices of the dodecahedron on the Bloch sphere. Given that Alice is not allowed to report null-results, it holds that $\mathcal{W}_0\leq (3+\sqrt{5})/10$ for LHS models \cite{Saunders2010} and  $\eta_\text{crit}\approx 0.227$ \cite{Vallone2013}. While for perfect measurements this is significantly lower than the efficiency needed for the ESI, the situation is the opposite when operating at $\epsilon=\epsilon^*$: the LHS bound increases to $\mathcal{W}_{\epsilon^*}\lesssim 0.6081$ and the critical efficiency increases to $\eta_\text{crit}\approx 0.53$.

\subsection{The ($n=4$)-witness}\label{AppDetection}
We want to compute the critical detection efficiency, $\eta_{\text{crit}}$, required for Bob to be able to detect steering within the plateau length of the witness $\mathcal{W}_0$ in Eq. \eqref{witness}. When Alice's detector fails she deterministically assigns her output values in accordance with the first class of strategies in Eq.~\eqref{classes_four}. Additionally, we make the ansatz that the shared state between the parties is given by
\begin{equation}\label{eq:ansatz_state}
|\Psi\rangle_{AB} = \dfrac{\cos\theta}{\sqrt{2}}|00\rangle+\dfrac{\sin\theta}{\sqrt{2}}|11\rangle+\dfrac{\sin\theta}{\sqrt{2}}|22\rangle+\dfrac{\cos\theta}{\sqrt{2}}|33\rangle,
\end{equation}
valid for $\theta = [0, \pi/4]$. This yields that the value of the steering witness becomes $C_\theta = (2\cos^2\theta-1)$. The eigenvalue problem of finding the optimal quantum value $Q_\theta$ is then computed numerically. By solving $\eta Q_\theta + (1-\eta)C_\theta = 1$ for fix $\theta$, we obtain the optimal detection efficiency $\eta_{\text{crit}} = 1/4$ in the limit $\theta \rightarrow 0$. 

\begin{figure}[ht!]
	\centering
	\includegraphics[width=\columnwidth]{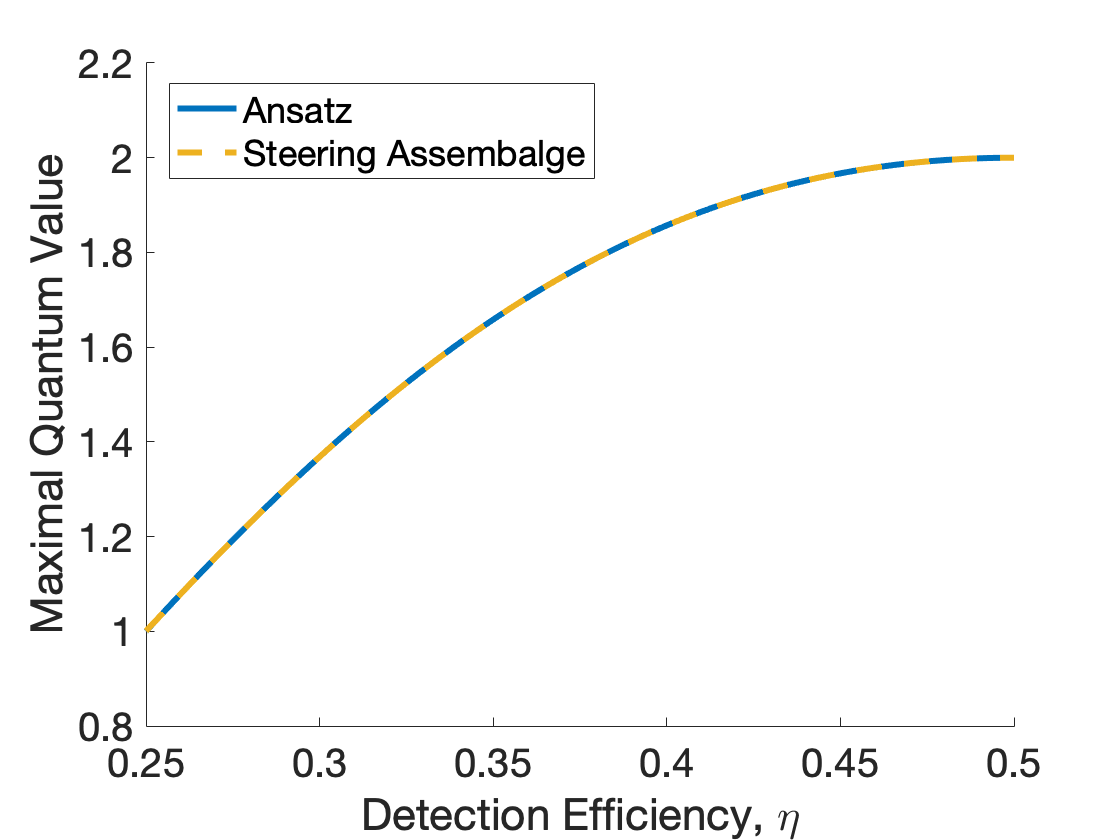}
	\caption{We compare $(Q_\theta, \eta_\theta)$ from ansatz \eqref{eq:ansatz_state} (blue solid curve) with $(Q, \eta)$ alternatively obtained via the SDP formulation in \eqref{eq:SDP_det} (yellow dashed curve).}\label{Fig_det_eff}
\end{figure}

In fact, the quantum value $Q_\theta$ obtained through the anstaz \eqref{eq:ansatz_state} is optimal. This can be verified by alternatively calculating this value over an arbitrary steering assemblage $\{\sigma_{a\mid x}\}$ for fix $\eta$. We formulate this optimisation task as an SDP 
\begin{equation}\label{eq:SDP_det}
\begin{aligned}
\max_{\{\sigma_{a\mid x}\}} \quad &Q = \sum_{abxy}c_{abxy}\operatorname{Tr}(B_{b\mid y}^{\text{targ}} \sigma_{a\mid x}) \\
\text{subject to} \quad &\eta Q + (1-\eta) C = 1\\
&\sigma_{a\mid x} \succeq 0 \quad \forall a,x\\
&\operatorname{Tr}(\sum_a \sigma_{a\mid x}) = 1 \quad \forall x \\
&\sum_a \sigma_{a\mid x} = \sum_a \sigma_{a \mid x'} \quad \forall x,x'
\end{aligned}
\end{equation}
where we express the steering value as $C = \operatorname{Tr}(B_1^{\text{targ}} \sum_a \sigma_{a\mid x})$. The optimal quantum values $Q_\theta$ and $Q$ are illustrated in Fig.~\ref{Fig_det_eff} as a function of $\eta_{\text{crit}}$. The former is obtained from the ansatz \eqref{eq:ansatz_state} (blue curve), whereas the latter is computed via \eqref{eq:SDP_det} (yellow curve). 

\section{Generalisation of theorem of Ref.~\cite{Tavakoli2024}}\label{AppGen}

We generalise the main theorem introduced in Ref.~\cite{Tavakoli2024} such that it applies to projective target measurements of rank greater than one when the trace of the lab measurement equals the rank of the target measurements.

Consider the the set of lab measurement $\{B_{b\mid y}\}$ in dimension $D$. The lab measurements are only constrained such that the fidelity of each POVM is limited by
\begin{equation}
\label{eq:fidelity_gen}
\operatorname{Tr}(B_{b\mid y} B_{b \mid y}^{\text{targ}}) \geq r(1-\epsilon_{b y})
\end{equation}
where $B_{b\mid y}^{\text{targ}} = \sum_{k = 1}^r |\psi_k\rangle \langle \psi_k|$ is the projective target measurement of rank $r$ and $\epsilon_{by}$ is the imprecision parameter associated to the $y$'th measurement with outcome $b$. Next, we define the semidefinite operator	
\begin{equation}
N = (1+\mu_{by}) B_{b\mid y}^{\text{targ}}+ \eta_{by} \mathds{1}_D
\end{equation}
for some parameters $\mu_{by} \geq -1$ and $\eta_{by} \geq 0$. The goal is to find parameters $(\mu_{by}, \eta_{by})$ such that $R \equiv \langle \chi | N-B_{b\mid y}| \chi \rangle \geq 0$ for every state $|\chi \rangle$ and measurement $B_{b\mid y}$ subject to ~\eqref{eq:fidelity_gen}. Thus, we start by decomposing $B_{b\mid y}$ into the eigenbasis of the target measurement. This is done as follows: First, let $B_{b\mid y} = \sum_i q_i |\phi_i\rangle \langle \phi_i|$ for some pure states $\phi_i$ and set $\{q_i\}$ such that $\sum_i q_i = \operatorname{Tr}(B_{b\mid y}) = r$, where $0 \leq q_i \leq 1$. Furthermore, each pure state in the ensemble can be decomposed into
\begin{equation}
\label{eq:psi_gen_th}
|\phi_i\rangle = \sqrt{1-\xi_i}|\tilde{\psi}_i\rangle + \sqrt{\xi_i}|\tilde{\psi}_i^{\perp}\rangle, \quad 0 \leq \xi_i \leq 1,
\end{equation}
where $|\tilde{\psi}_i^{\perp}\rangle$ is some state orthogonal to $|\tilde{\psi}_i\rangle \in \mathcal{H}_{targ}$. Next, we express $|\tilde{\psi}_i\rangle$ in the eigenbasis of $B_{b\mid y}^{\text{targ}}$, i.e., $|\tilde{\psi}_i\rangle = \sum_{j=1}^r z_j^{i}  |\psi_j\rangle$, where $\sum_j |z_j^{i}|^2 = 1$ to preserve normalization of the state. In particular, we let $z_j^{i} = e^{i\alpha_{j}^{i}}y_j^{i}$, where $\alpha_{j}^{i}$ is taken to be the relative phase between $z_j^{i}$ and the first coordinate $z_1^{i}$, and $y_j^{i}$ is the $j$th coordinate of the unit $n$-sphere $(n = r)$. Explicitly, the coefficients are given by
\begin{equation}
\begin{aligned}
&z_1^{i} = \cos\theta_1^{i} \\
&z_j^{i} = e^{i\alpha_{j}^{i}} \cos\theta_j^i \prod_{k=1}^{j-1}\sin\theta_k^{i}  \quad j = 2,...,r-2\\
&z_{r-1}^{i} = e^{i\alpha_{r-1}^{i}} \sin\varphi^{i}\prod_{k=1}^{r-2}\sin\theta_k^{i}  \\
& z_r^{i} = e^{i\alpha_{r}^{i}} \cos\varphi^{i} \prod_{k=1}^{r-2}\sin\theta_k^{i} 
\end{aligned}
\end{equation}
where the angles are defined for $\theta_j^{i} \in [0,\pi]$, $\varphi^{i} \in [0, 2\pi)$, and $\alpha_j^{i} \in [0, 2\pi)$. Note that $\alpha_1^{i} = 0$. We observe that the fidelity condition ~\eqref{eq:fidelity_gen} now can be written as
\begin{equation}
\sum_i q_i \xi_i \leq r\epsilon_{by}.
\end{equation}
Now, to determine the parameters $(\mu_{by}, \eta_{by})$ we expand $R$
\begin{equation}
R = \eta_{by} + (1+\mu_{by}) \sum_{k=1}^r |\langle \chi| \psi_k\rangle|^2 - \sum_i q_i |\langle \chi | \phi_i \rangle |^2.
\end{equation}
Specifically, the third term can be further expanded to 
\begin{equation}
\begin{aligned}
|\langle \chi | \phi_i \rangle |^2 &= (1-\xi_i) \sum_j (y_j^{i})^2 |\langle \chi|\psi_j\rangle|^2 \\
&+\xi_i |\langle \chi | \tilde{\psi}_i^{\perp} \rangle|^2\\
&+ 2(1-\xi_i)\sum_{j < k} y_j^{i} y_k^{i} \text{Re}(e^{i(\alpha_j^{i}-\alpha_k^{i})}\langle \chi | \psi_j\rangle \langle \psi_k | \chi \rangle) \\
& + 2\sqrt{\xi_i(1-\xi_i)}\sum_j y_j^{i} \text{Re}(e^{i\alpha_j^{i}}\langle\chi|\psi_j\rangle \langle \tilde{\psi}_i^{\perp} | \chi\rangle) )
\end{aligned}
\end{equation}
At this point, we introduce the variables $x_j = \langle \chi|\psi_j \rangle = |x_j|e^{i\tilde{\phi}_j}$ and $x_i^{\perp} = \langle \chi | \tilde{\psi}_i^{\perp}\rangle =  | x_i^{\perp}|e^{i\tilde{\phi}_i^{\perp}}$, where $\sqrt{\sum_j |x_j|^2 + |x_i^{\perp}|^2} \leq 1$, due to conservation of probability. Since we require that $R \geq 0$ for every choice of $(x_1,...,x_r, x_i^{\perp})$ we can choose the phases such that $|\langle \chi | \phi_i \rangle |^2$ is maximised. This implies that $\tilde{\phi}_1 = ... = \tilde{\phi}_r = \tilde{\phi}_i^{\perp}$ and that the relative phase $\alpha_j^{i} = 0$ for all $j$. Thus, we can without loss of generality proceed with $(x_1,...,x_r,x_i^{\perp})$ being real parameters with the same sign, meaning that we constrain the angles to $\theta_j^{i} \in [0,\pi/2]$ and $\varphi^{i} \in [0,\pi/2]$. As a result,
\begin{equation}
|\langle \chi | \phi_i \rangle |^2  \leq \sum_i q_i (\sqrt{1-\xi_i} \sum_j y_j x_j + \sqrt{\xi_i}x_i^{\perp})^2
\end{equation}
 From Cauchy-Schwartz inequality we now get that $\sum_j y_j x_j = \vec{y} \cdot \vec{x} \leq ||\vec{y} || || \vec{x} || = || \vec{x} || $, since $||\vec{y} || = 1$. Denoting $x = || \vec{x} || = \sqrt{\sum_j x_j^2}$ and observing that $R$ is minimised for $x_i^{\perp} = \sqrt{1-x^2}$, we obtain
\begin{equation}
\begin{aligned}
R \geq &\eta_{by}-\sum_i q_i \xi_i+x^2\left(\mu_{by}+2 \sum_i q_i \xi_i\right)-\\
&2 x \sqrt{1-x^2} \sum_i q_i \sqrt{\xi_i\left(1-\xi_i\right)}
\end{aligned}
\end{equation}
Thereafter, the rest of the argument follows directly from Ref.~\cite{Tavakoli2024}, which gives
\begin{equation}\label{eq:B_by}
\begin{aligned}
B_{b\mid y} &\preceq(1+\mu_{by}) B_{b\mid y}^{\text{targ}}\\
&+\frac{1}{2}\bigg(\sqrt{\mu_{by}^2+4 (r\epsilon_{b y})(1+\mu_{by})}-\mu_{by} \bigg) \mathds{1}_D\\
&= (1+\mu_{by}) B_{b\mid y}^{\text{targ}} +\frac{1}{2}z_{by} \mathds{1}_D.
\end{aligned}
\end{equation}
Furthermore, $z_{by}$ is non-negative and monotonically decreasing in $\mu_{by}$. Therefore the optimal choice of parameters that minimse the right hand side of the inequality in $\eqref{eq:B_by}$ is to take all $\mu_{by}$ as large as possible, setting $\mu = \max_{by} \mu_{by}$. In this case, the $\vec{\epsilon}$-dependent LHS bound of an arbitrary witness is bounded by
\begin{equation}\label{eq:W_eps_bound}
\begin{aligned}
&\mathcal{W}_{\vec{\epsilon}} \leq \min_{\mu \geq -1} (1+\mu)\beta_0 + \dfrac{1}{2}\sum_x \max_a \big(\sum_{by} c_{abxy} z_{by}\big)
\end{aligned}
\end{equation}
where $\beta_0$ is the standard LHS bound for perfect measurements and 
\begin{equation}
z_{by} = \sqrt{\mu^2+4 (r\epsilon_{b y})(1+\mu)}-\mu.
\end{equation}
Note that the result \eqref{eq:W_eps_bound} is general and that every choice of the parameter $\mu \geq -1$ corresponds to a valid analytical bound.

\end{document}